\documentclass{aa}
\usepackage{graphicx}
\usepackage{txfonts}
\usepackage{array}
\usepackage{graphics}
\usepackage{latexsym}
\usepackage{amssymb}
\usepackage{amsmath}
\usepackage{fancyhdr}
\usepackage{morefloats}
\usepackage{rotating}
\bibpunct{(}{)}{;}{a}{}{,} 
\begin{document}

   \title{The parallelism between galaxy clusters and early-type galaxies:}

   \subtitle{III. The Mass-Radius Relationship}

   \author{C. Chiosi \inst{1},
          M. D'Onofrio\inst{1}\fnmsep\thanks{Corresponding author}, E. Merlin \inst{2}, L. Piovan \inst{1} \and P. Marziani \inst{3}
          }

   \institute{Department of Physics and Astronomy, University of Padova, Vicolo Osservatorio 3, I35122, Italy\\
              \email{cesare.chiosi@unipd.it, mauro.donofrio@unipd.it, lorenzo.piovan@gmail.com}
         \and
             INAF - Astronomical Observatory of Rome, Via Frascati 33, I00078 Monte Porzio Catone (RM), Italy\\
             \email{emiliano.merlin@inaf.it}
         \and
             INAF - Astronomical Observatory of Padua, Vicolo Osservatorio 5, I35122, Italy\\
             \email{paola.marziani@inaf.it}
%
             }

   \date{Received June 20, 2020; accepted July 20, 2020}

   \titlerunning{The Mass-Radius Relationship}
   \authorrunning{Chiosi, D'Onofrio, Merlin, Piovan, Marziani}


  \abstract
  {This is the third  study of a series dedicated to the observed parallelism of properties between Galaxy 
Clusters and Groups(GCGs) and    early-type galaxies (ETGs).}
   {Here we investigate the physical origin of  the Mass-Radius Relation (MRR). }
  {Having collected literature data on masses and radii for objects going from Globular Clusters (GCs) to 
ETGs and GCGs,  we set up the  MR-plane and compare the observed distribution with the  MRR  predicted by  theoretical 
models both for the monolithic and  hierarchical  scenarios. }
 { We argue that the distributions of stellar systems in the MR-plane is due to complementary mechanisms: 
 (i) on one hand, as shown in paper II, the relation of the virial equilibrium does intersect with a relation 
 that provides the total luminosity as a function of the star formation history; (ii)  on the other hand, the 
 locus predicted for the collapse of systems should be convolved with the statistical expectation for the maximum 
 mass of the halos at each comsic epoch. This second aspect provides a natural boundary limit explaining either the 
 curved distribution observed in the MR-plane and the existence of a zone of avoidance.}
   { The distribution of stellar systems in the MR-plane is the result of two combined evolution, that of the 
   stellar component and that of the halo component.  }


   \maketitle
%

\section{Introduction}\label{sec:intro}

In recent years, much attention has been paid to the Mass-Radius Relationship (MRR) of galaxies,
 the early-type galaxies (ETGs) in particular. The MRR is indeed basic to any theory of
galaxy formation and evolution. To this aim, an impressive body of data have been acquired concerning 
the masses and  dimensions of galaxies and galaxy clusters and groups not only in the local old Universe 
but also in the distant and young one, thus making it feasible to address
the question whether these two important parameters changed with time as predicted by any hierarchical mode
of galaxy formation, see for instance the studies by
\citet[][]{Shankar_etal_2011},
\citet{Bernardi_etal_2011},  \citet{Graham2011, Graham2013}, \citet{Bernardi_etal_2014},
\citet{Agertz_Kravtsov_2016}, \citet{Kuchner_etal_2017},
\citet{Huang_Fall_2017},
\citet{Somerville_etal_2018}, \citet{Genel_etal_2018}, \citet{Terrazas_etal_2019},  and references
therein.

The subject of the MRR of galaxies from ETGs to dwarf ellipticals and dwarf
spheroidals (dwarf galaxies, DGs, in general), including also bulges and Globular Clusters (GCs) has been
reviewed by \citet{Graham2011, Graham2013} to
whom we refer for many details.  The current MRRs for ETGs and DGs will
be presented below. At the same time  data for the mass and radius of Galaxy Clusters and
Groups (GCGs) have been acquired \citep[see for instance ][WINGS data, and references
therein]{Valentinuzzi_etal_2011, Cariddi_etal_2018}. Therefore, it is worth of interest to investigate whether 
a similar MRR  exists for this class of objects and how it would compare with the one of galaxies.

In addition to this, convincing evidence has been gathered that at relatively high redshifts, objects
of mass comparable to that of nearby massive galaxies but with smaller dimensions exist. These ``compact
galaxies'' are found up to $z \geq 3$
with stellar masses from $10^{10}$ to $10^{12} M_\odot$ and half-light radii from 0.4 to 5 kpc
(i.e. 3 to 4 times  more compact than the local counterparts of the same mass), and in nearly similar
proportions there are galaxies with the
same mass but a variety of dimensions \citep[e.g.,][]{Mancini_etal_2009,Valentinuzzi_etal_2010a},
 and bulge to disk ratios \citep[e.g.,][]{Vanderwel2011}.
However, here we want to start considering only the case of standard ETGs and GCGs, leaving the  compact
galaxies aside.

In \cite{Donofrio_etal_2019b} (hereafter paper II) we have already analyzed the origin of the scaling relations 
from the point of view of the stellar component. We have demonstrated that these relations originate from the combination 
of the virial theorem with the luminosity evolution of the galaxies.
The aim of this study is instead that of looking at the general physical principles governing in particular the MRR.

We  will address in particular the question: why do GCGs, ETGs and GCs obey a rather narrow MRR instead of
scattering around showing a broader combination of these two parameters? 
Similar analyses have been
made by \citet{Chiosi_Carraro_2002} and \citet{Chiosi_etal_2012},
so that the present study is a sequel of those ones motivated by the better data nowadays at our disposal.  To
clarify the aims and the methods of the present study (and also the previous ones), we anticipate
here the essence of the analysis. We speculate that the observed MRR   is the result of two
complementary mechanisms: (i) 
Objects from GCs  to ETGs and  GCGs 
crowd on the MR-plane along loci with the same slope but  different zero points. We name these loci 
model MRRs (M-MRRs). They are parameterized by the initial density (redshift) at which these proto-systems collapse.
The slopes of the M-MRRs are about a factor of two flatter than the observational MRR.
The M-MRRs are scarcely affected by the dominant formation mechanism at work, either quasi-monolithic or hierarchical. Furthermore,
along each mass sequence, the evolutionary history of the massive galaxies tend to reach  the ideal case of the dissipation-less  
collapse, whereas  the less massive ones significantly depart from this. 
(ii)   
At each redshift the  mass distribution
 galactic halos  is set by cosmology, i.e. there is the upper mass limit for collapsed objects. This implies  a 
limit to the mass extension of each M-MRR of the manifold, i.e. a boundary line  on the MR-plane, the slope of which 
changes with the 
mass (roughly more than a factor of two on logarithmic scale for masses and radii). The galaxies along this 
boundary  are   in mechanical and thermal equilibrium. 
If a galaxy does not meet these conditions, its position on the MR-plane is above the boundary line. 
The intersection of the M-MRR manifold  with the boundary lines at different redshift yields the observed 
MRR for objects going from GCs to GCGs.

The paper is subdivided as follows. In Section \ref{Obs_MRR} we present and discuss the observational MRR for
three samples of galaxies and galaxy clusters.
In Section \ref{models_illustris} we shortly describe the hydrodynamical models of galaxies (and clusters) over a
large range of masses contained in the large scale simulations by
\citep{Vogelsberger_2014a,Vogelsberger_2014b,Genel_etal_2014,Nelson_etal_2015}
that are adopted here as the main source of theoretical data\footnote{The more recent data of Illustris-TNG
have not been included in the present study to secure homogeneity with the previous ones of the series.}.
In Section  \ref{models_illustris} we derive the theoretical  MRR from the large scale simulations both for
the
present day Universe and as function of the redshift and try to highlight the main causes determining the
observational MRR.  In Section \ref{Theo_MRR} based on elementary
theories of cosmology and galaxy formation we derive the same MRR and compare it with that obtained from the
large scale simulation showing that good agreement exists. In Section \ref{MRR_cosmo} we seek to derive  the MRR
from First Principles highlighting the deep causes that eventually determine its shape.
Finally, in Section \ref{conclusions} we draw some remarks and conclusions.

Throughout the paper we assumed in all our calculations the same values of the $\Lambda$-CDM cosmology used
in the Illustris simulations by \citet{Vogelsberger_2014a,Vogelsberger_2014b}:
$\Omega_m = 0.2726, \Omega_{\Lambda}= 0.7274, \Omega_b = 0.0456, \sigma_8 = 0.809, n_s = 0.963, H_0 = 70.4\,
km\, s^{-1}\, Mpc^{-1}$.

\section{The observational Mass-Radius Relation} \label{Obs_MRR}

In this section we present the data for single galaxies and galaxy clusters and groups.
These are: (i) the catalog of ETGs,  spiral galaxies, DGs, GCs  and GCGs  in the Local Group and local
Universe by   \citet{Burstein_etal_1997};  (ii)  the SDSS data for ETGs by
\citet{Bernardi_etal_2010} roughly covering the redshift interval z=0 to $\simeq 2$;
(iii) the  samples  of ETGs, Bright Central Galaxies (BCGs)  and GCGs set up
by \citet{Valentinuzzi_etal_2011, Cariddi_etal_2018} using the data  of the WINGS survey;
(iv) finally, the list of dwarf galaxies
for the local Group is supplemented by that of \citet{Woo2008}, \citet{Geha_etal_2006},
\citet{Hamraz_etal_2019}, and that of galactic GCs of \citet{Burstein_etal_1997} by the data of
\citet{Pasquato_Bertin_2008} and the transition objects from GCs to DGs of \citet{Kissler-Patig_etal_2006}.

\begin{figure*}
\centering{
\includegraphics[width=0.85\textwidth]{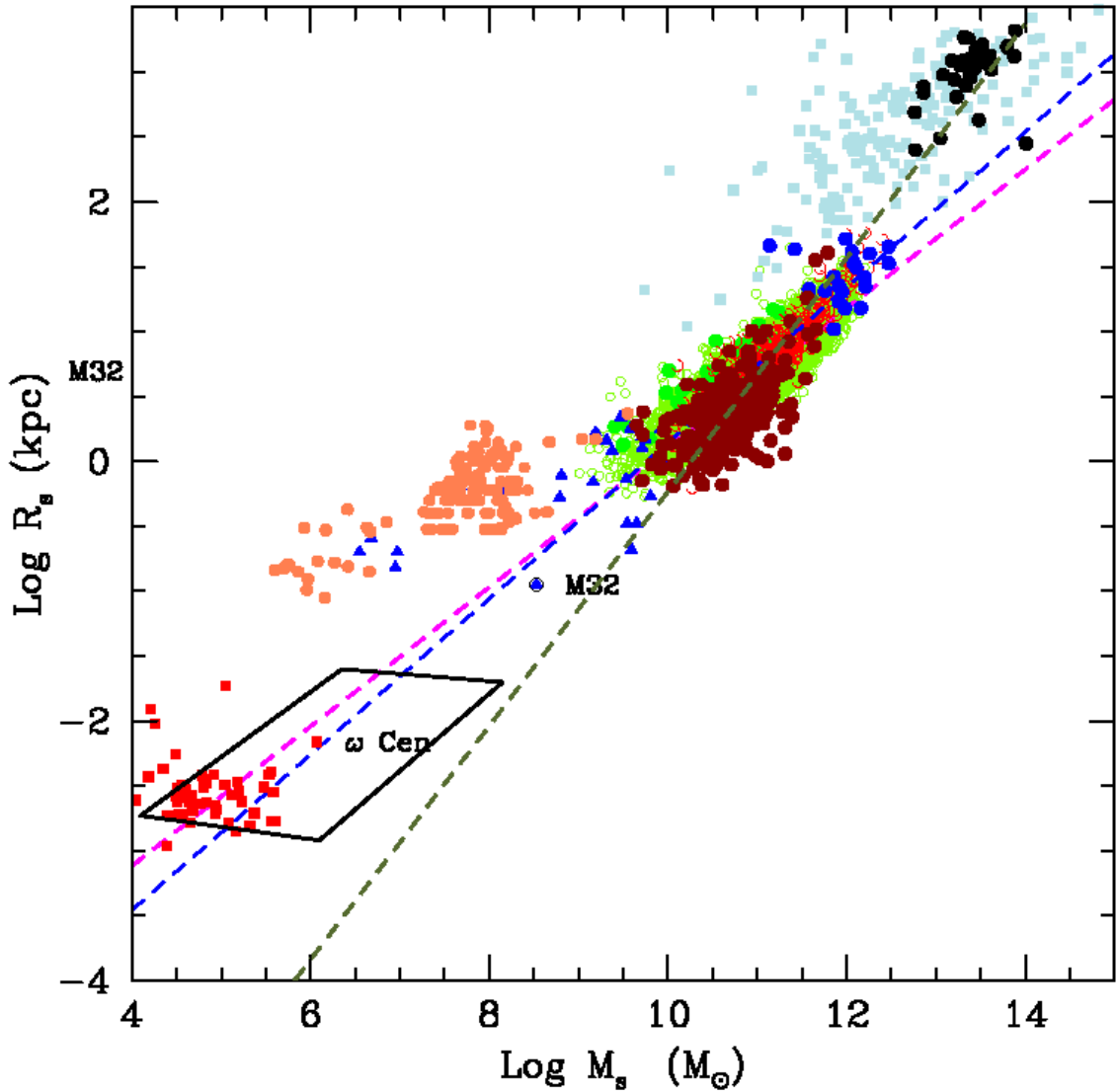}  }
\caption{ The $\log (R_{s})$ versus $\log (M_{s})$ relation for  all the
samples under consideration. $R_{s}$ is the half-mass radius, and
$ M_{s}$ the total stellar mass. Although $R_{s}$ is not strictly identical to the effective radius
$R_e$,
they are very close to each other. Throughout this paper we will always use $R_{s}$, which is easier to
calculate for hydrodynamical models of galaxies, and assume $R_{s}\simeq R_e$.
\textsf{\citet{Burstein_etal_1997} sample}:
the filled red  circles are the ETGs, the green open circles the SGs, the blue filled triangles the DGs, the filled
red squares the GCs,  the filled light-blue squares the GCGs. The
blue dashed line is the linear best fit of eqn. (\ref{RsMs_Burstein}),
i.e. $\log R_{s}=0.60\,\, \log M_{s} - 5.859$ relative to the sole ETGs, however extended
 to the regions of GCs  and GCGs.
\textsf{\citet{Bernardi_etal_2010} sample}:  only ETGs are present indicated by green filled circles.
The dashed  magenta line is the linear best fit of the data extended to GCs and GCGs.
\textsf{WINGS sample}: the dark-red filled circles  are the ETGs, the blue filled circles  the
BCGs, and the black  filled circles the GCGs. The olive-green dashed line is the linear best-fit of all
WINGS objects together $\log (R_{s} =
0.901 \log M_{s} -9.245$, however extended to DGs and GCs.
\textsf{Dwarf galaxies and Globular Clusters}: the coral filled circles are
the DGs of \citet{Woo2008} and \citet{Geha_etal_2006} all together. Finally, the parallelogram shows the
area occupied by the transition
objects from GCs to DGs of \citet{Kissler-Patig_etal_2006}.
}
\label{all_data}
\end{figure*}

\textsf{The Burstein et al. (1997) data}.  Over the years these very popular data have been examined by many
authors so that a detailed presentation is superfluous here. We limit ourselves to show in Fig. \ref{all_data} 
the distribution on the $\log R_s$ vs $\log M_s$ plane of the various subgroups of the
data set from the classical GCs  to GCGs: the red filled circles are the
ETGs,  the green open  circles the spiral galaxies crowding the same area but not considered in the analysis, 
the blue triangles the dwarf galaxies, the red squares  the
globular clusters, finally the light-blue filled squares the ETG-rich and Sp-rich Clusters/Groups  all together. 
  The  best fit of the sole ETGs yields for $M_s \geq 10^{10} \, M_\odot$

\begin{equation}
\log R_s = (0.59 \pm 0.01) \log M_s - (5.85 \pm 0.16) 
\label{RsMs_Burstein}
\end{equation}

\noindent where  $M_{s}$ (in $M_\odot$) is the estimated stellar mass
 and $R_{s}$ (in kpc) is the radius
containing half of it (nearly identical to the classical effective radius $R_e$), $rms=0.13$ and the 
correlation parameter $corr= 0.91$. This relation is taken from
\citet{Chiosi_Carraro_2002} who used the same data. 
This line is then extended to  the domain of GCs and of GCGs, however  arbitrarily shifted by -0.1
dex on both coordinates,  to highlight the role of this line as a border  of the
observed distribution of astrophysical objects whose mass extends over about eleven orders of magnitude
(the blue dashed line).

\textsf{The ETGs of \citet{Bernardi_etal_2010}}. A much richer sample of data for ETGs has been derived
by \citet{Bernardi_etal_2010} from the SDSS catalog. The sample contains $\simeq 60,000$ galaxies
\footnote{The selection
conditions are $\textrm{fracDeV}=1$ and $b/a> 0.6$, therefore the sample is dominated by elliptical galaxies 
(see \citet{Bernardi_etal_2010} for more details).} The observational MRR is displayed in  Fig. 
\ref{all_data}, 
the green filled circles. The linear best fit of the SDSS data is

\begin{equation}
\log R_{s}=(0.537 \pm 0.001)\,\,\log M_{s} - (5.26 \pm 0.01)
 \label{RsMs_Bernardi}
\end{equation}

\noindent where  $M_{s}$  and $R_{s}$ have their usual meaning and units (the magenta dashed line), $rms= 
0.094$ and the correlation parameter $corr= 0.89$. 
The slope (and zero point) of the above
MRR is quite robust as nearly  coincides with similar determinations made by other authors: e.g.
 \citet{Chiosi_Carraro_2002} using the \citet{Burstein_etal_1997} data and \citet{Shenetal2003} using the
SDSS
data. We will show that the same slope is also recovered using the Illustris simulations.

The distribution of the bulk of galaxies is confirmed by the smaller sample of
\citet[][]{Shankar_etal_2011} also
extracted from the SDSS survey but using slightly different selection criteria. The area covered by the
observational data
is slightly larger than the one with the \citet{Bernardi_etal_2010} data.

\textsf{The WINGS database}.
In recent times, large optical and spectroscopic databases for the galaxy content of nearby clusters
have become available  thanks to the WIde-field Nearby Galaxy-cluster Survey (WINGS) of
\citet{Fasano_etal_2006} and \citet{Varela_etal_2009} and companion OMEGA-WINGS extension of
\citet{Gullieuszik_etal_2015} and \citet{Moretti_etal_2017} for a number of clusters in the
redshift interval ($0.04 \leq z \leq 0.07$).
All this material has been subsequently examined by
\citet{Cariddi_etal_2018} with particular attention to the problems of the accurate
determination  of the stellar light  and the
stellar mass profiles of galaxy clusters.  
{WINGS measured and examined thousand of galaxies in 46 clusters, 
providing the absolute V and B magnitudes,  the  morphological types (according to the
classification system  RC3;  
\citet{deVaucouleurs_etal_1991}, \citet{Corvin_etal_1994} and
\citet{Fasano_etal_2012}), four different estimates of the star formation rates (SFR) and the estimates of the stellar masses $M_s$ 
\citep{Fritz_etal_2007,Fritz_etal_2011} and the effective radii $R_e$ and effective surface brightness \citep{Donofrio_etal_2014}}.
The issue of the membership of the galaxies to the clusters
under consideration has been addressed
and examined by \citet{Cava_etal_2009} to whom we refer for all details.
In this study we have considered all the 46 clusters studied
by \citet{Cariddi_etal_2018}.  The MR-plane of this set of data is shown in Fig. \ref{all_data} where the dark-red filled 
circles are the ETGs, the blue filled circles the BCGs, and the black filled circles the GCGs.
The inspection of the WINGS data reveals that: (i) compared to \citet{Burstein_etal_1997} and
\citet{Bernardi_etal_2010} at given mass the radii of ETGs are smaller by about 0.3 dex whereas  those of BCGs
and GCGs are comparable; (ii)
the MRR relation for the ETGs more massive than $10^{10} \, M_\odot$ is

\begin{equation}
\log R_{s}=(0.46 \pm 0.04)\,\,\log M_{s} - (4.60 \pm 0.40)
\label{mr_wings1}
\end{equation}

\noindent where masses and radii are in the usual units, $rms=0.22$ and $cor=0.61$;
(iii) the scatter around this reference MRR   is much larger than
in the previous cases; (iv) ETGs and Spirals crowd in the same region of the MR-plane;
(v) extended plumes at the large mass side of the MRR may exist only for
ETGs   belonging to clusters and not for ETGs belonging to the field and in general not for both field
and cluster late type galaxies \citep{Donofrio_etal_2019b}; (vi) looking at
the cluster ETGs, the MRR has curved banana-like shape with a well developed plume toward high masses and
radii made of red galaxies as confirmed by their  B-V color  and  also their S\'ersic index n
\citep[][]{Donofrio_etal_2019a}. In contrast, in field ETGs the banana-like structure of the MRR and
the red plume is much less
evident if not missing at all. The reason for this striking difference is not clear, most likely it is
related to the higher probability for massive cluster ETGs of merging and/or engulfing other galaxies of
smaller mass in the case of wet mergers to avoid any bluing effect in their colors or of comparable mass
and age in case of dry mergers among similar objects thus leaving the color unchanged \citep[see the
discussion in ][]{Sciarratta_etal_2019}.

Considering all the objects together (ETGs, BCGs and GCGs) the MMR is 
\begin{equation}
\log R_s = (0.90 \pm 0.01) \log M_s - (9.24\pm 0.18) 
\label{wings_all}
\end{equation}

\noindent with $rms=0.28$ and $cor=0.95$ (the dark olive-green dashed line in Fig. \ref{all_data}). It is worth noting
that adopting eq. (\ref{wings_all}) as the actual MRR for galaxies and galaxy clusters and basing on its slope, one might  
conclude that the observed MRR simply mirrors the slope of 1 expected from the   virial equilibrium condition
\citep{Donofrio_etal_2019b}. 
In this 
study we will show  that in reality
the MRR owns its origin to a more complicate interplay among different causes.

\textsf{Dwarf Galaxies}. The DGs are taken from different sources: (i) the
\citet{Burstein_etal_1997} sample of dEs and dSphs of the Local Group (indicated as B-DGs).
It is worth recalling that the masses used by \citet{Burstein_etal_1997} are the dynamical
masses and not the stellar masses,
so this group is not strictly homogeneous with the sample for ETGs;   (iii) the sample of DGs by
\citet{Geha_etal_2006} (indicated as G-DGs); (iii) the
DGs of the Local
Group according to the measurements made by \citet{Woo2008} (indicated as W-DGs). 
Fortunately, all the three samples of data   yield  much similar MRRs

\begin{eqnarray}
\log R_{s}&=&(0.22 \pm 0.05)\log M_{s} -(2.1\pm 0.5)\label{dwarfB} \\
\log R_{s}&=&(0.22 \pm 0.07)\log M_{s} -(1.9\pm 0.5)\label{dwarfG} \\
\log R_{s}&=&(0.28 \pm 0.03)\log M_{s} -(2.4\pm 0.2)\label{dwarfW}  
\end{eqnarray}

\noindent Eq.(\ref{dwarfB}) is for the B-DGs  with $rms=0.281$ and $cor=0.632$, eq.(\ref{dwarfG}) for the G-DGs 
with $rms=0.193$ and $cor=0.300$, and Eq.(\ref{dwarfW}) fot the W-DGs
$rms=0.201$ and $cor=0.851$. 
Here we   consider the three relationships as identical, but give more weight  to that of  \citet{Woo2008} 
with the highest 
correlation parameter. Finally, we take into account the study
by \citet{Kissler-Patig_etal_2006} on the
transition  objects from GCs to dwarfs galaxies. All these data are shown in
Fig. \ref{all_data}.

\textsf{Galaxy Clusters and Groups}.
Two sources of data for galaxy clusters and groups  have been considered, namely  \citet{Burstein_etal_1997}
and   the WINGS and Omega-WINGS database \citep{Fasano_etal_2006, Varela_etal_2009, Cava_etal_2009,
Moretti_etal_2014, Donofrio_etal_2014, Gullieuszik_etal_2015, Moretti_etal_2017,
Biviano_etal_2017, Cariddi_etal_2018}. In particular the parameters $R_s$ and $M_s$  needed to the present
study are those measured by  \citet{Biviano_etal_2017} and  \citet{Cariddi_etal_2018}. See also
\citet{Donofrio_etal_2019a} for more details.

\textsf{On the MRR slope}. 
It is soon evident that there is no unique slope for the MRR of 
ETGs and DGs. The slope for ETGs goes from 0.5 to 0.6, and for dwarf galaxies from
0.217 to 0.272. Furthermore,
looking at the data in detail, the slope is even steeper than 0.54 in the region of the largest and most
massive ETGs going up to 1 and even more, see the top part of the MRR by \citet[][]{Bernardi_etal_2010},
\citet{Guo2009}, \citet{vanDokkum2010}, \citet[Fig. 1 in ][]{Graham2011} and \citet{Graham2013}.
This is a point to keep in mind when interpreting the observational data.

\textsf{General Remarks}.
Information and details on how the stellar masses $M_{s}$ and half-mass radii, $R_{s}$, have been derived can be found 
in the original sources to which the reader should refer. Of course some possible systematic biases
among the different sets of data are to be expected, whose entity, however, ought to be small. This is
somewhat sustained by the overall agreement among different sources as far as some general  relationships are
concerned, e.g.  the agreement in the slope of the MRR for ETGs between \citet{Bernardi_etal_2010} and
\citet{Burstein_etal_1997}. The same for the dwarf galaxies.  However, since the  groups of objects will be
treated separately and only from a general qualitative point of view, no homogenization of the data is
needed. Furthermore, despite the  important remarks about the WINGS galaxies, we can say that there is no
substantial difference passing from the MRR based on the \citet{Burstein_etal_1997} data, to the one based
on the \citet{Bernardi_etal_2010} data, and finally the WINGS data. Our analysis of the MRR for ETGs (and
partially spirals as well) will primarily stand on the SDSS sample of \citet{Bernardi_etal_2010}, thus
securing internal homogeneity of the mass and radius estimates. Finally, in this study
the MRR derived from the \citet{Bernardi_etal_2010} data
will be considered as the reference case.

\section{Current theoretical  models of galaxies and galaxy clusters} \label{models_illustris}

Our aim here is to present the numerical simulations of galaxies and clusters that we have used to interpret
and reproduce the observed properties of real galaxies and clusters. 

The primary source of data is the Illustris
compilation\footnote{http://www.illustris-project.org/data/} \citep[][to whom we refer for all
details]{Vogelsberger_2014a,Vogelsberger_2014b, Genel_etal_2014,Nelson_etal_2015}, a suite of large,
highly detailed
cosmological hydrodynamical simulations, including star, galaxy and black-hole formation and tracking the expansion of the
universe \citep{Hinshaw_etal_2013}.
The procedure we have adopted to extract theoretical data from the Illustris database is amply described in
\citet{Donofrio_etal_2019b}  to whom the reader should refer for all details.
Suffice to mention here that in order to follow the evolution of each galaxy we have extracted from the
Illustris database the data of
stellar mass, dark matter mass, total mass,  luminosity, half-mass radius of the stellar component, velocity
dispersion and star formation rate for the whole
set of galaxies (with mass $\log(M_s)\geq 9$ at $z=0$) in the selected clusters at the redshift $z=0$,
$z=0.2$, $z=0.6$, $z=1$, $z=1.6$, $z=2.2$, $z=3$ and $z=4$. With these data we have analyzed the MRR   at
different epochs following the progenitors of each object.

In this section we present a quick analysis of the Illustris sample of  galaxy models.
Hereinafter $M_D$, $R_D$, $M_B$, $R_B$,
$M_s$  and $R_s$  are masses and half-mass radii of Dark Matter (DM), Baryonic Matter (BM) made of
stars and gas (initially only gas) and stellar mass (initially zero), respectively. The total mass of a
galaxy is defined as $M_T= M_D+M_B$.    At the beginning  the ratio $M_D / M_B = \omega$ is fixed by the cosmological
model of the Universe, in our case for $\Omega_m/\Omega_b$ $\omega= \simeq 5.92\simeq 6$. 
It is worth keeping in mind that in the course of
the formation and evolution processes the above masses can change in presence of galactic winds and/or
stripping and/or acquisition of  material by interactions with other galaxies or intergalactic medium.

\begin{figure*}
\centerline{
\includegraphics[width=7.0cm,height=7.0cm]{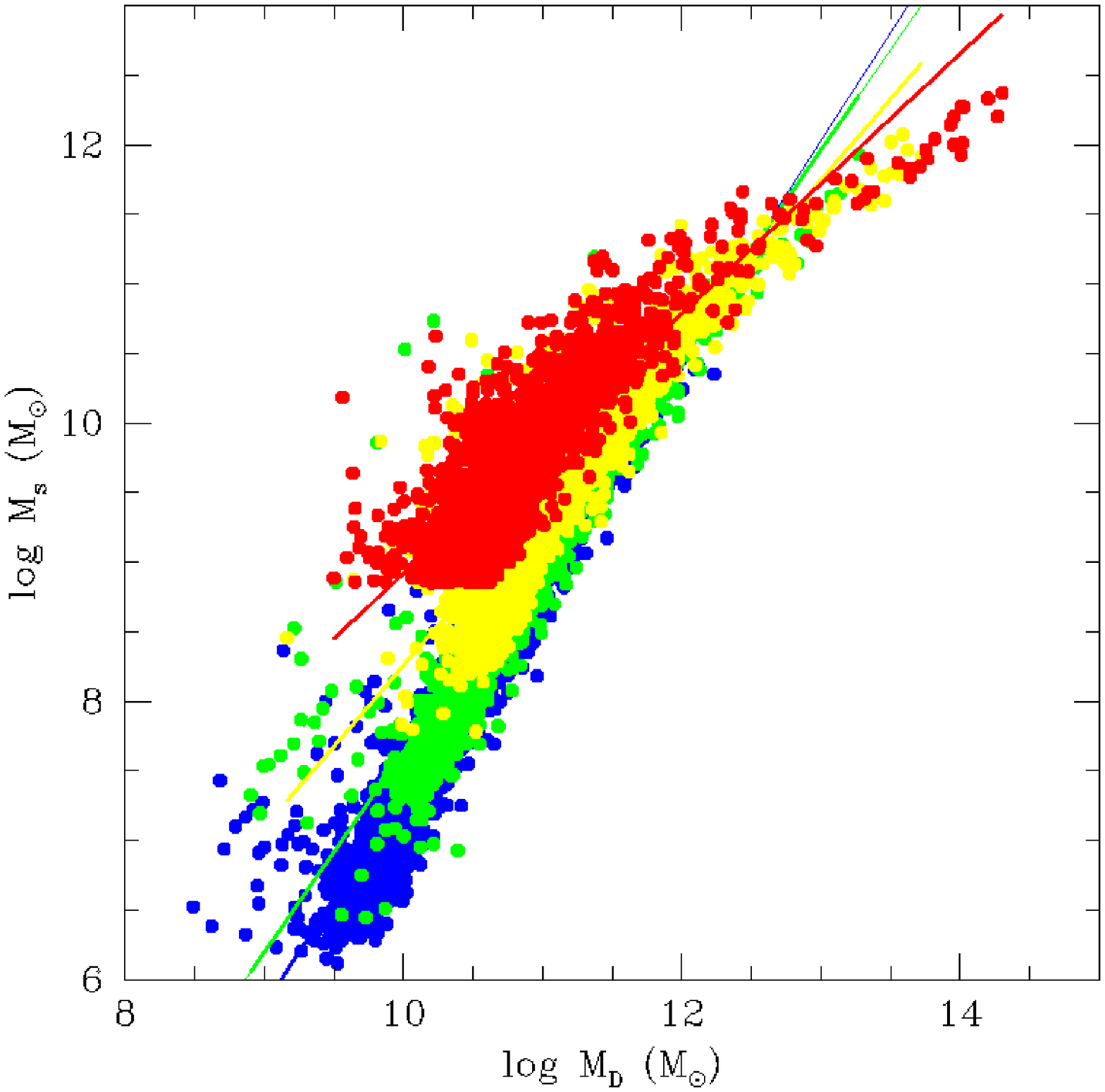}
\includegraphics[width=7.0cm,height=7.0cm]{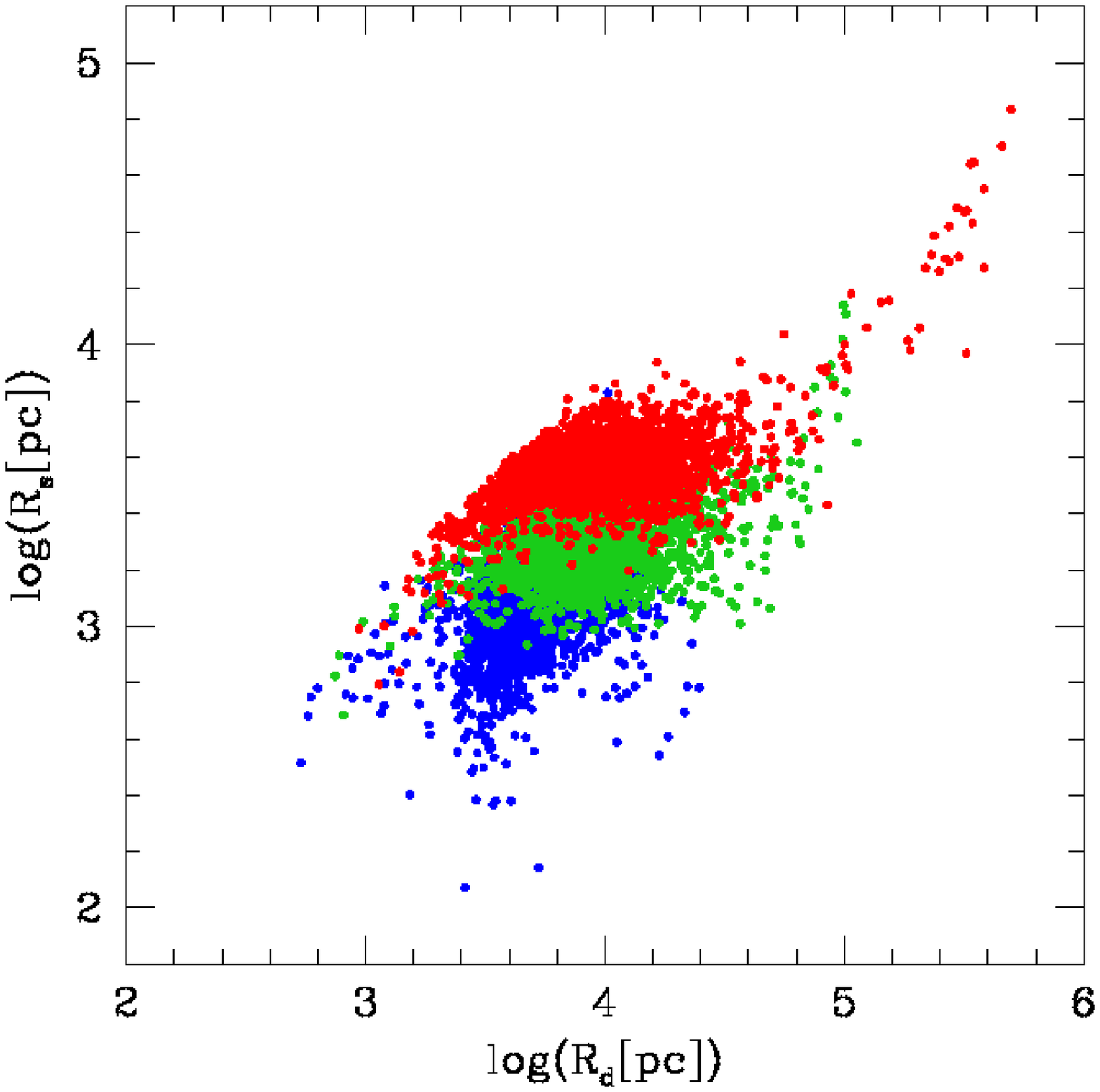}  }
\caption{ \textbf{Left Panel}: The $M_s - M_D $ relations at different redshifts (z=4, blue; z=2, green; z=1,
yellow;  z=0,  red). Masses are in solar units. The solid  lines are the best fits, the coefficients of which
are given in Table \ref{ms_md_rs_rd_tab}.  \textbf{Right Panel}: the same as in the left panel but for
the $R_s - R_D$ relations. Radii are in kpc.}
\label{ms_md_rs_rd}
\end{figure*}

\subsection{ Stars and DM: relationships among $M_s$, $M_D$, $R_s$ and $R_D$ }
The relationships   $M_s$ vs $M_D$ and $R_s$
vs $R_D$ for four different values of the redshift (z= 4, 2, 1, and 0) are shown in the left and right panels of 
Fig. \ref{ms_md_rs_rd}, respectively. Masses (in $M_\odot$) and radii (kpc) are in logarithmic notation and the color 
code indicates the redshift (z=4, blue; z=2, green; z=1, yellow; z=0, red).

\textsf{The $M_s$ vs $M_D$ relationship}. At the beginning of the galaxy formation
history, DM ad BM are in cosmological proportions (i.e. $M_{D} = \omega M_{B}$ with $\omega \simeq 6$), but star 
formation gradually stores
more and more baryonic mass into stars. It may be worth of interest to evaluate the efficiency of the star
formation process over the Hubble time  in galaxies of different mass.
Since $M_B < M_D$ so  $M_s$ is always smaller than $M_D$. However, since galaxies of different total mass may build 
stars at different efficiencies, the ratio $M_s$ to $M_D$ is expected not to be constant, but
to vary with $M_D$ and redshift. In the left panel of Fig. \ref{ms_md_rs_rd}, we note that $M_s$ always increases with $M_D$ 
so that low mass galaxies build up less stars with respect to the more massive ones, however the slope of 
the relation as the redshift goes to zero. In more detail,
for redshift s $\left(z\gtrsim 2\right)$ and masses 
$M_D \simeq 10^{12}\, M_\odot$ the slope
decreases at decreasing redshift so that more and more stars are present at given $M_D$. More precisely, for 
$\left(z\lesssim 2\right)$ and  $M_D \leq 10^{12}\, M_\odot$ the above trend holds good to, but above this mass limit the 
opposite seems to occurs, at given $M_D$ less star mass is present than expected. In other words, massive galaxies are less 
efficient builders of their stellar content (the star formation is over).
The relationship between $\log M_s$  and $\log M_D$ is  $M_s=\alpha M_D + \beta$ and the best-fits are given  
in  Table \ref{ms_md_rs_rd_tab}.

To quantify the efficiency of star formation we calculate the ratio $M_s/M_D$ as a function of $M_D$. In this case we can 
neglect the change of slope in the $M_s$ vs $M_D$ relation above a certain value of $M_D$ at low redshifts. Finally
Finally we get the inverse of $M_s/M_D$ given by the expression

\begin{equation}
 \frac{M_D}{M_{s}} = 10^{-\beta} M_{D}^{1-\alpha}  \qquad  {\rm and } \qquad  \frac{M_D}{M_{s}} = \omega \frac {M_B}{M_s}
\label{inverse_md_ms}
\end{equation}
where the ratio $M_B/M_s$ measures how much of the original BM mass is turned into stars.

\noindent
The  values of ($1-\alpha$) and $-\beta$ are listed in Table \ref{eff_coeff} for the  four redshifts we are considering.

For the sake of illustration we list in Table \ref{efficiency} the ratio $M_s/M_D$ as a function of the total
mass  limited to the case of $z=0$.
In general the ratios $M_s/M_D$ and $M_s/M_B$ decrease with the total galaxy mass.
Similar results were found
by  \citet{Chiosi_Carraro_2002}, \citet{Merlin2006,Merlin2007} and \citet{Merlin2012} thus showing 
that old models were already able to catch the essence of the galaxy formation problem.

\textsf{The $R_s$ vs $R_D$ relationship}. In similar way we derive the relations $R_s = \eta R_{D}^{\gamma}$ that 
are shown in the right panel of
Fig. \ref{ms_md_rs_rd} and the entries of Table \ref{ms_md_rs_rd_tab} for the case of $z=0$. The radius of
$R_D$ is  more extended than $R_s$ by a factor of about 3 to 10 as the galaxy mass
increases from $10^9 \, M_\odot$ to $10^{13}\, M_\odot$. 
The slope $\gamma$ of $R_s - R_D$ relation
(logarithmic) first decreases by  about a factor of two passing from z=4 to z=1, and then increases again at
z=0.  What is more important is that while at high redshifts (our z=4, z=2 and z=1 cases) the galaxy
distribution on the $R_s$ vs $R_D$ plane is a random cloud of points, at z=0 a regular trend gets in
place
in which $R_s$ increases with $R_D$ on the side of large values of $R_D$ (largest masses). On the side of low
values of both  radii and masses a cloud of points is still there. The effect of this is to increase the mean slope of
the
whole distribution. What does it imply? We will  try to cast light on this issue.

\begin{table}
\begin{center}
\caption{The $M_s$ vs $M_D$ and $R_s$ vs $R_D$ relations. For $M_s$ vs $M_D$ the analytical form is
adopted:  $\log M_s = \alpha \log M_D + \beta$ ($M_s=10^{\beta}M_{D}^{\alpha}$). For
$R_s$ vs $R_D$  the expression $\log R_s = \gamma \log R_{D}+  \eta$ ($R_s=10^{\eta}R_{D}^{\gamma}$) is used.  
 Masses and radii are in $M_\odot$ and kpc, respectively. The models are from the Illustris catalog.   }
\label{ms_md_rs_rd_tab}
\begin{tabular}{|c | c| c|   c|| c| c| c|}
\hline
\multicolumn{1}{|c|}{$z$}              &
\multicolumn{3}{c||}{ $M_s$ vs $M_D$ } &
\multicolumn{3}{c|}{  $R_s$ vs $R_D$ } \\
\hline
         & $\alpha$     & $\beta $  &  $M_D$       &  $\gamma$ & $\eta$& $R_D$  \\
\hline
 4       & 1.55         & -8.19     &              &  0.39     &-0.20      & \\
\hline
 2       & 1.44         & -6.78     &              &  0.30     & 0.03      &  \\
\hline
 1       & 1.16         & -3.37     &  $ < 12.0$   &  0.22     & 0.07      & \\
         & 0.76         &  2.30     &  $ > 12.0$   &  0.22     & 0.07      & \\
\hline
0.       & 0.93         & -0.43     &  $ < 11.5$   &  0.296    & -0.087    &\\
         & 0.79         &  1.22     &  $ > 11.5$   &  0.294    & -0.042    &\\
\hline
\end{tabular}
\end{center}
\end{table}

\begin{table}
\begin{center}
\caption{Coefficients of the relationships  $ \log \left(M_D/Ms\right) =(1- \alpha)\,\log M_D  - \beta$ with at
different redshifts.
The data are taken from  the Illustris catalog of model galaxies. Masses  are in $M_\odot$.  }
		\label{eff_coeff}
		\begin{tabular}{|c | l| l| }
\hline
\multicolumn{1}{|c|}{Redshift}&
\multicolumn{1}{c|}{$\alpha$ } &
\multicolumn{1}{c|}{$-\beta$ }  \\
\hline
4.0   & -0.555   & 8.186 \\
3.0   & -0.498   & 7.514 \\
2.0   & -0.441   & 6.782 \\
1.0   & -0.100   & 2.927 \\
 0    &  0.093   & 0.229 \\
\hline
\end{tabular}
\end{center}
\end{table}

\begin{table}
\begin{center}
\caption{Efficiency of the star formation  in galaxies of different mass observed at redshift $z=0$. The
efficiency is measured by the ratio $M_s/M_{D}$. The data are taken from  the Illustris catalog
of model galaxies. Masses  are in $M_\odot$.  }
		\label{efficiency}
		\begin{tabular}{ | l| l| l| c| c| c|}
\hline
\multicolumn{6}{|c|}{Redshift z=0} \\
\hline
\multicolumn{1}{|l|}{$M_{D}$ } &
\multicolumn{1}{c|}{${M_{B}}$ } &
\multicolumn{1}{l|}{$M_{T}$} &
\multicolumn{1}{l|}{$\log M_s$} &
\multicolumn{1}{c|}{$\frac{M_s}{M_{D}}$ } &
\multicolumn{1}{c|}{$\frac{M_D}{M_{s}}$ }  \\
\hline
 $10^{12}$  & $1.67\, 10^{11}$ & $1.167\, 10^{12}$ & 10.770  &0.058 & 17.12\\
 $10^{11}$  & $1.67\, 10^{10}$ & $1.167\, 10^{11}$ &  9.863  &0.073 & 13.71\\
 $10^{10}$  & $1.67\, 10^{9}$  & $1.167\, 10^{10}$ &  8.852  &0.089 & 11.17\\
 $10^{9}$   & $1.67\, 10^{8}$  & $1.167\, 10^{9}$  &  8.057  &0.114 &  8.77\\
 $10^{8}$   & $1.67\, 10^{7}$  & $1.167\, 10^{8}$  &  7.162  &0.145 &  6.88\\
 $10^{7}$   & $1.67\, 10^{6}$  & $1.167\, 10^{7}$  &  6.267  &0.182 &  5.48\\
\hline
\end{tabular}
\end{center}
\end{table}

\begin{figure}
\centerline{
\includegraphics[width=7.0cm,height=7.0cm]{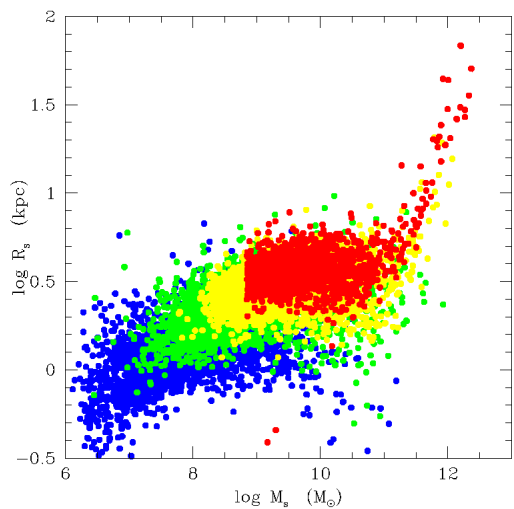} }
\caption{ The stellar  half-mass radius $R_s$ vs the mass  $M_s$ of the galaxy models of the Illustris
database at different values of the
redshift, i.e.  z=4 (blue), z=2 (green), z=1 (yellow) and z=0 (red).   }
\label{mass_radius_4z}
\end{figure}

\subsection{The $R_s$ vs $M_s$ relation at different redshifts}

In Fig. \ref{mass_radius_4z} we show the $R_s$ vs $M_s$ relations of the Illustris models at the four values of
the redshift considering (the color code we have adopted.
Like  the case  of  $R_s$ vs $R_D$ relations, the distribution is clumpy and irregular at high redshifts independently 
of the galaxy mass. Starting from z=1 and more evident at  z=0 a tail-like feature develops at the side of
large masses, say $\gtrsim 1-2\cdot10^{11}\, M_\odot$.

The best fit of the data  at redshift z=0 using the relationship $R_s = \eta M_s^{\epsilon}$ (where masses
 and radii are
in $M_\odot$ and kpc, respectively) yields the values listed in Table \ref{ms_rs_tab}. At higher redshifts,
the tail at the side of large masses is much less evident if not missing at all. The tail seems to disappear starting from
z=1. The $R_s$ vs $M_s$ relation is much similar to that 
of the low mass galaxies at z=0, i.e. nearly
flat. At any value of the mass in the mass range of the data cloud, the dispersion in the radius is very large. At
redshift z=0, the tail  has slope and zero point much similar to those derived
by  \citet{Chiosi_etal_2012}  using the SDSS data of \citet{Bernardi_etal_2010}.

What is the reason for the cloud-like and tail-like distributions at low redshifts? Why the cloud-like one
dominates in the low mass range and at high redshifts? The opposite trend happens for the tail-like one, which shows up 
in the high mass range and at low redshifts.   What is the
physical meaning of the two distributions? To cast light on these issues we examine the history of the  
$R_s$ vs $M_s$ relation  of a number of individual galaxies.

\begin{figure}
\centerline{
\includegraphics[width=7.0cm,height=7.0cm]{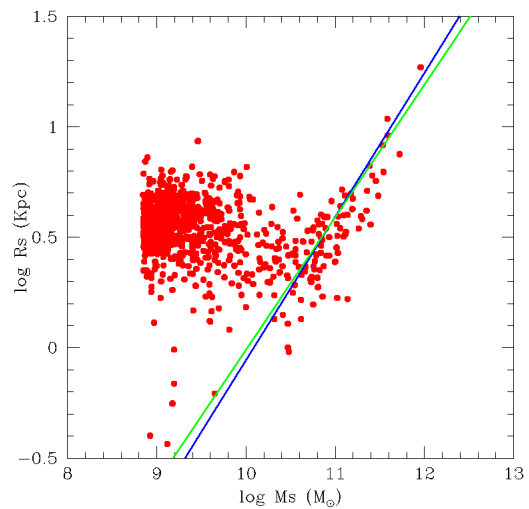}  }
\caption{ The same of Fig. \ref{mass_radius_4z} but limited to the sample for z=0 for all models with null
star formation rate. }
\label{mass_radius_z0_sfr0}
\end{figure}

\begin{table}
\begin{center}
\caption{The radius-mass relation of the stellar component $R_s = \eta M_s^{\epsilon}$. Masses and radii are
in $M_\odot$ and kpc, respectively. Each relation is a two slopes one with separation mass $M_s= 2\times 10^{11}
M_\odot$
suggested by the data. }
		\label{ms_rs_tab}
		\begin{tabular}{|c | c| c| c| c|}
\hline
\multicolumn{1}{|c|}{Case}&
\multicolumn{2}{c|}{$M_s \leq 2\times 10^{11}\, M_\odot$ } &
\multicolumn{2}{c|}{$M_s >    2\times 10^{11}\, M_\odot $} \\
\hline
  & $\epsilon$   &   $\eta$  & $\epsilon$    &   $\eta$     \\
\hline
 Illustris               &-0.005  &  0.592 & 0.651 & -6.557 \\
 Chiosi et al. (2012)    & 0.187  & -1.626 & 0.570 & -5.694\\
\hline
\end{tabular}
\end{center}
\end{table}

\subsection{The detailed history of the $R_s$ vs $M_s$ relation  for  selected models}

The Illustris simulations are based on the hierarchical scheme, therefore each galaxy is the result of
a number of
mass acquisition/removal processes, which change the masses $M_D$, $M_s$ and the radii $R_D$ and $R_s$ of the galaxy. 
For each galaxy in the sample at z=0, the Illustris data base provides the past history, i.e. the
masses and radii of the components sub-units during the Hubble time. This means that we can reconstruct the
past history in the $R_s$ vs $M_s$ plane of each galaxy from z=4 to z=0. 
The path of each galaxy on the MR-plane is quite tortuous:
We can summarize summarize the complex situation as follows: in mergers
among low mass objects  the mass and the radius increase, however, exceptions are possible,
where either the mass or the radius or both. In
general the model galaxies remain inside the cloud-like region of the $R_s$ vs $M_s$ plane.  Mergers among galaxies
of relatively high mass tend to generate objects that shift outside the cloud and tend to fall close to  a
well behaved radius-mass sequence (actually they define it) and their locus agrees with the observational
radius-mass relationship for ETGs \citep[see e.g.][and references therein]{Chiosi_etal_2012}.  Finally, the
cloud-like region coincides with the distribution of dwarf galaxies of different type \citep[see the
discussion by][]{Chiosi_Carraro_2002}. The MRR for massive galaxies is very close to the
relation 
set by the condition of virial equilibrium (this issue will be examined in great detail below), so one might be
tempted to conclude that systems that at the present time (z=0) that are able to satisfy the virial condition
have the minimum energy and hence radius for their existing structure. Dwarf galaxies, most likely because
they are undergoing active star formation, cannot be in this ideal condition. So the question arises
spontaneously: do dwarf galaxies exist that fulfill the virial state? What determines the large radius of dwarf
galaxies well visible in Fig. \ref{all_data}. Recasting the question in a different way: are there DGs whose position 
on the MR-plane is near  the  MRR of normal ETGs? In other words in conditions of mechanical and thermal equilibrium. The
following considerations can be made.

Given the mass of a galaxy (either acquired by mergers or already in place ''ab initio``), the radius
mirrors the condition of mechanical equilibrium of the system. In other words  it is a consequence of the
energy balance between external dynamical processes (collapse) or internal feed-back by star formation and
other
sources. To answer the above question one should look at galaxies in which at least star formation
activity has extinguished since a reasonable amount of time.
To this aim we consider the z=0 sample (most likely containing  many objects with null star formation)
and isolate the galaxies with null star formation rate (SFR=0).  These are plotted in  Fig. \ref{mass_radius_z0_sfr0} 
the galaxies with SFR=0 at $z=0$. 
In addition to galaxies of high mass, there are also a few objects of low mass that fall very
close to the prolongation of best-fit line of the  \citet{Bernardi_etal_2010} ETGs
 \citep[see][]{Chiosi_etal_2012}. Their observational counterparts could be objects like $\omega$Cen and M32.
 So it seems that also low mass galaxies can exist sharing the same equilibrium conditions 
 of the massive ones. This conclusion is however biased by the large uncertainties on the SFR,
which is known
only up to three decimal digits. A SFR of the order of $10^{-4} \, M_\odot /yr$ or lower cannot be neglected
in the case of dwarf galaxies. This may explain why even for this case a residual cloud-like feature still
remains.

\subsection{Comparison between observations and theory}\label{compare_theor_obser}

\begin{figure*}
\centerline{
{\includegraphics[width=0.85\textwidth]{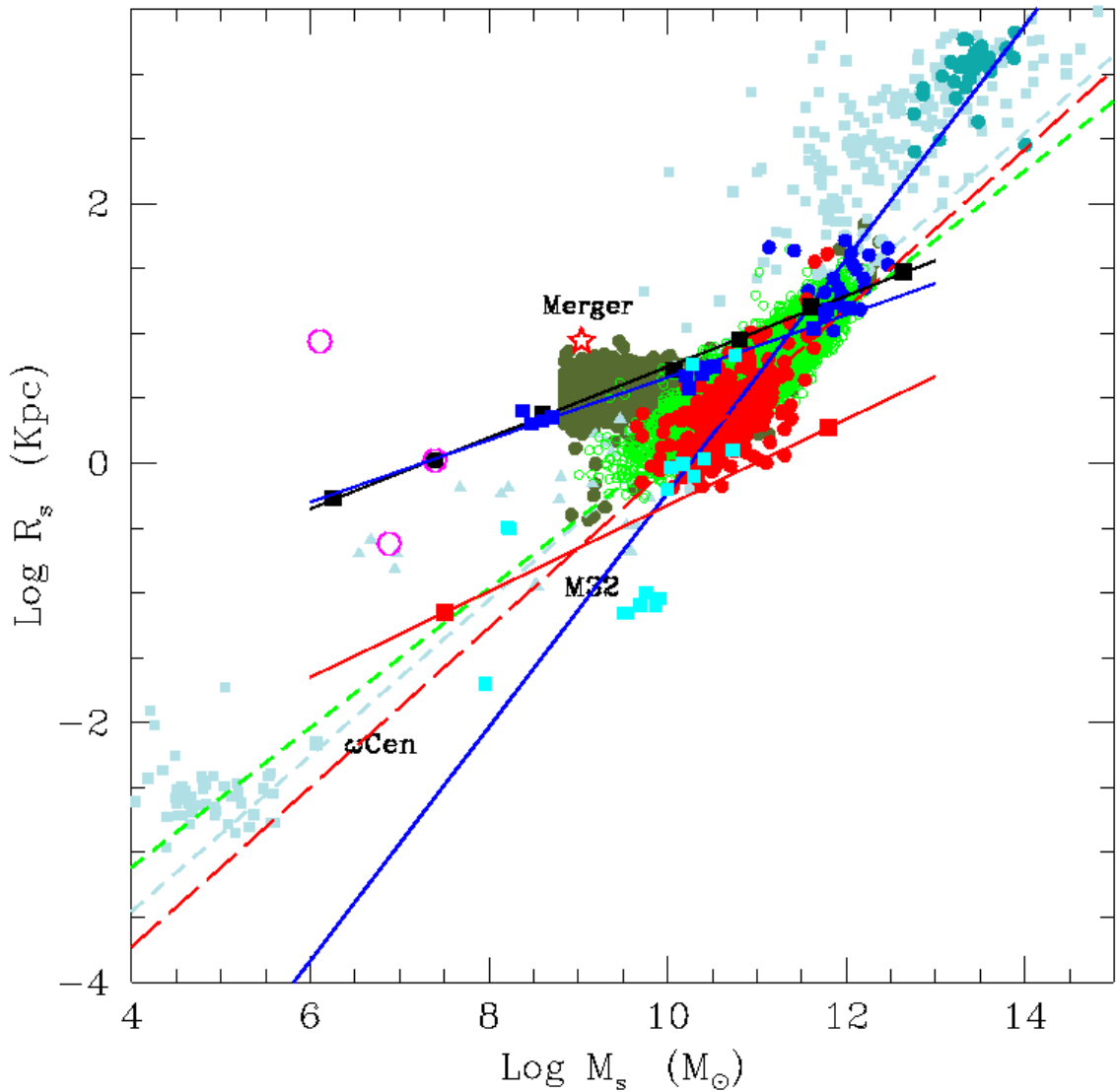} }   }
\caption{ The composite $R_s$ vs $M_s$ relation and comparison between theory and observations. 
The light powder-blue dots are the data of \citet{Burstein_etal_1997} for ETGs, DGs, GCs and GCGs: the  dashed line of the same colour
is the best fit of the sole ETGs however extended to the domains of
GCs and GCGs. The bright green dots are the \citet{Bernardi_etal_2010} data for ETGs, and  the  thick dashed line 
of the same colour is  the best-fit line of these latter. The red dots, blue dots, and green-blue tots   are the ETGs, BCGs, and
GCGs of the WINGS sample. The dashed red line
is the best-fit line of the  ETGs, whereas the solid blue line is  the best fit of all WINGS data lumped together. 
The slope of this latter line is very close to  MRR for virialized
objects. In such case the line fails to hit the region of Globular Clusters. Finally, we display  different theoretical 
models for comparison. The black filled squares and
the black solid line are the monolithic models by \citet{Chiosi_Carraro_2002}  and their linear fit. 
The three magenta open circles are low mass ETGs with the same mass but different initial densities by 
\citet{Chiosi_Carraro_2002}.See the data of
Table \ref{tab1}. The empty red star is the case of merger reported by \citet{Chiosi_Carraro_2002}. 
The blue squares and the blue line are the hierarchical models by \citet{Merlin2010,Merlin2012}, see Table \ref{tab2}. 
The light blue-green squares are the ancillary model of
\citet{Chiosi_etal_2012}, see Table \ref{tab2}. The dark live-green points are the reference model galaxies of the 
Illustris simulations. }
\label{composite_RsMs}
\end{figure*}

Before proceeding further, it is mandatory to compare  the observational data with
theoretical galaxy models in usage. This is shown in Fig. \ref{composite_RsMs} which displays the
following data:
i) \citet{Burstein_etal_1997} sample of ETGs, DGs, GCGs, and GCs that is a fairly homogeneous set of
sample.  
ii) \citet{Bernardi_etal_2010} of ETGs. This sample is by far more numerous than the previous one 
for galaxies of the same type.  iii). The WINGS data for ETGs, BCGs and GCGs  by \citet{Valentinuzzi_etal_2011,
Cariddi_etal_2018}. The emphasis here is given to the group of ETGs.  The theoretical models are those of
the hydrodynamical large scale simulations of the Illustris
project \citet{Vogelsberger_2014a, Vogelsberger_2014b} at redshift $z=0$ (the olive green dots) in the pure hierarchical scheme,
those by 
\citep[][]{Chiosi_Carraro_2002}  in the pure monolithic scenario, and finally those by 
\citep{Merlin2006,Merlin2007,Merlin2010,Merlin2012, Chiosi_etal_2012} in the so-called early-hierarchical view, 
both at redshift $z=0$. The emphasis is given to the Illustris models that are considered as the reference case.
From this comparison we see that

i) The \citet{Burstein_etal_1997} data for ETGs  fairly agree  with the Illustris models of comparable mass. Unfortunately 
Illustris does not extend enough
into the regions populated by DGs, simply because for technical reasons the sample at $z=0$ is limited
in mass at $10^9 \, M_\odot$. In addition to it, the observational sample to disposal contains too few
DGs. Therefore, nothing can be said for this type of objects.  
As already note
the MRR of ETGs  given by eq. (\ref{RsMs_Burstein}) has  slope $\approxeq 0.6$, even though the most massive ETGs 
would be better represented by a slope $\simeq 1$ thus suggesting a MRR  the slope of which slightly changes at increasing 
mass. For the moment we leave it aside. We will come back to it later on. This best-fit-line for ETGs,  extended
downward to the domain of GCs  and
upward to that of GCGs and vertically shifted by $\Delta \log R_s \simeq 0.3$, would match  all three groups
of objects  and leave all the data in the semi-plane above it.

The same
considerations and results are derived from using the \citet{Bernardi_etal_2010}, for which the best fit MRR 
of the sole ETGs is given by eq.(\ref{RsMs_Bernardi}),
which  is only slightly different from the previous one. They agree with the \citet{Burstein_etal_1997} data
and the Illustris models. The extension of the ETGs' MRR  also hit the GCs and GCG.

The WINGS data all together suggest a steeper slope of
the MRR, i.e. 0.95$\pm$0.02. The inclusion of galaxy clusters has
forced the slope to higher values.  Extending this relation (the solid line) to the domain of GCs would
not match these objects.
Furthermore, the WINGS data do not cover the region of galaxies with   $M_s
\simeq 10^9 \, M_\odot$ so that they do not completely overlap the area reached by Illustris models.
However, if we derive the MRR limited to the case of ETGs, the agreement between theory (Illustris) and data
is remarkable and the best-fit line of the ETGs alone (dashed line) would hit the region of GCs.

The main conclusion of this mutual comparison between data from different sources and the theoretical
models of Illustris is that all of them seem to fairly agree each other. As a matter of facts,
different sources of observational data, different photometry, and different volume coverage of the
space, but similar results. This is very important, because they suggest that the conclusions are
not severely affected by the source of data in usage.

Finally, we proceed to compare theoretical models with other theoretical models. The black filled squares
connected by the black line are the models by \citet{Chiosi_Carraro_2002} for low initial over-density
with respect to the surrounding medium and different mass, whereas the red squares connected by the
red line are models
of the same type but different mass and very high initial over-density contrast. The three coral
circles  are monolithic models of the same mass ($10^9\, M_\odot$) but different initial density contrast
by \citet{Chiosi_Carraro_2002}. The blue filled squares connected by the blue line are the models by  
\citet{Merlin2010,Merlin2012}. The cyan filled squares are
incomplete models of different mass and initial over-density limited to the very first  evolutionary stages
calculated by \citet{Chiosi_etal_2012} according to the early hierarchical scheme. They were meant
to localize the initial position of model galaxies on the MR-plane.
Details on the input/output parameters of all the model galaxies are given in Table \ref{tab0}, \ref{tab1},
and  \ref{tab2} of
Appendix \ref{Appendix_C}. It is worth noting
that the slope of the MRR of the models at  varying the mass but keeping constant the initial conditions
(over-density) are very similar each other. 

Remarkably, there is substantial agreement among the various types of models. Taking the Illustris case
as a reference, the monolithic and early hierarchical models fall onto the same position on the MR-plane, the
only difference being due to the richness of the three samples. While the Illustris models amount to more
than 2500 cases of different mass (total and stellar) and initial conditions that are picked up from 
large scale simulations \citep[see][for more details]{Vogelsberger_2014a, Vogelsberger_2014b}, those by
\citet[][]{Chiosi_Carraro_2002} and \citet{Merlin2006,Merlin2007,Merlin2010,Merlin2012, Chiosi_etal_2012}
are much fewer in number and different way of defining the initial conditions.  The
\citet[][]{Chiosi_Carraro_2002}  models were designed and calculated one by one assuming the
initial over-density of the proto-cloud with respect to
the surrounding cosmological medium and the initial positions and velocities of the DM and BM particles,
Those by \citet{Merlin2006,Merlin2007,Merlin2010,Merlin2012, Chiosi_etal_2012} stem from mini-large-scale
numerical cosmological simulations (about 10 Mpc by 10 Mpc)
that allowed for repeated mergers among sub-clumps of DM and BM in the same field. In this respect they
are somewhat similar
to the models by  \citet[][]{Vogelsberger_2014a, Vogelsberger_2014b}. From this comparison, we also learn  another important fact:
at given total stellar mass $M_s$ at the present time ($z=0$) galaxies tend to have the same size $R_s$ independently of the past 
formation history (pure monolithic, early hierarchical, pure hierarchical) because they share the same area of the MR-plane.

Furthermore, hydrodynamical simulations of  galaxies that  more or less
share the same initial conditions tend to converge to the same locus on the MR-plane  no matter
of details in input physics and numerical technique. In contrast, models with the same total mass but
different initial conditions converge to different loci on the MR-plane that however run parallel each other.
 Higher initial density objects tend to remain
smaller in size during their whole life. Finally, from this comparison among different galaxy  models we learn
that also those obtained with modest computing resources and simplified physical input and  numerical technique 
are fully adequate to explore a large variety of astrophysical problems.

The most important issues and questions raised by this section are (1) understanding the physical meaning of
the line splitting the MR-plane in two regions: (a) the one containing the observational data
for objects with mass
spanning about ten orders of magnitude from GCs to GCGs; (b) the one void
of objects with exception of the much fewer compact galaxies \citep[see][for a short discussion of
the issue]{Chiosi_etal_2012}; (2) the fact that this line is unique
and the separation is very sharp; (3) in principle galaxies of suitable mass and/or initial density could
fall in the ''forbidden semi-plane'' but for some yet not clear reasons the vast majority of real galaxies do
not. It is worth recalling that the ``forbidden semi-plane'' coincides with the region named ``Zone of
avoidance (ZOE) by \citet{Burstein_etal_1997}. A plausible explanation of the ''forbidden semiplane`` has
been advanced long ago by  \citet{Chiosi_Carraro_2002} and
\citet{Chiosi_etal_2012}. Considering the recent wealthy of modern data and theoretical models,  in the
following we go over it again.

\begin{figure}
\centering{
\includegraphics[width=8.0cm,height=8.0cm]{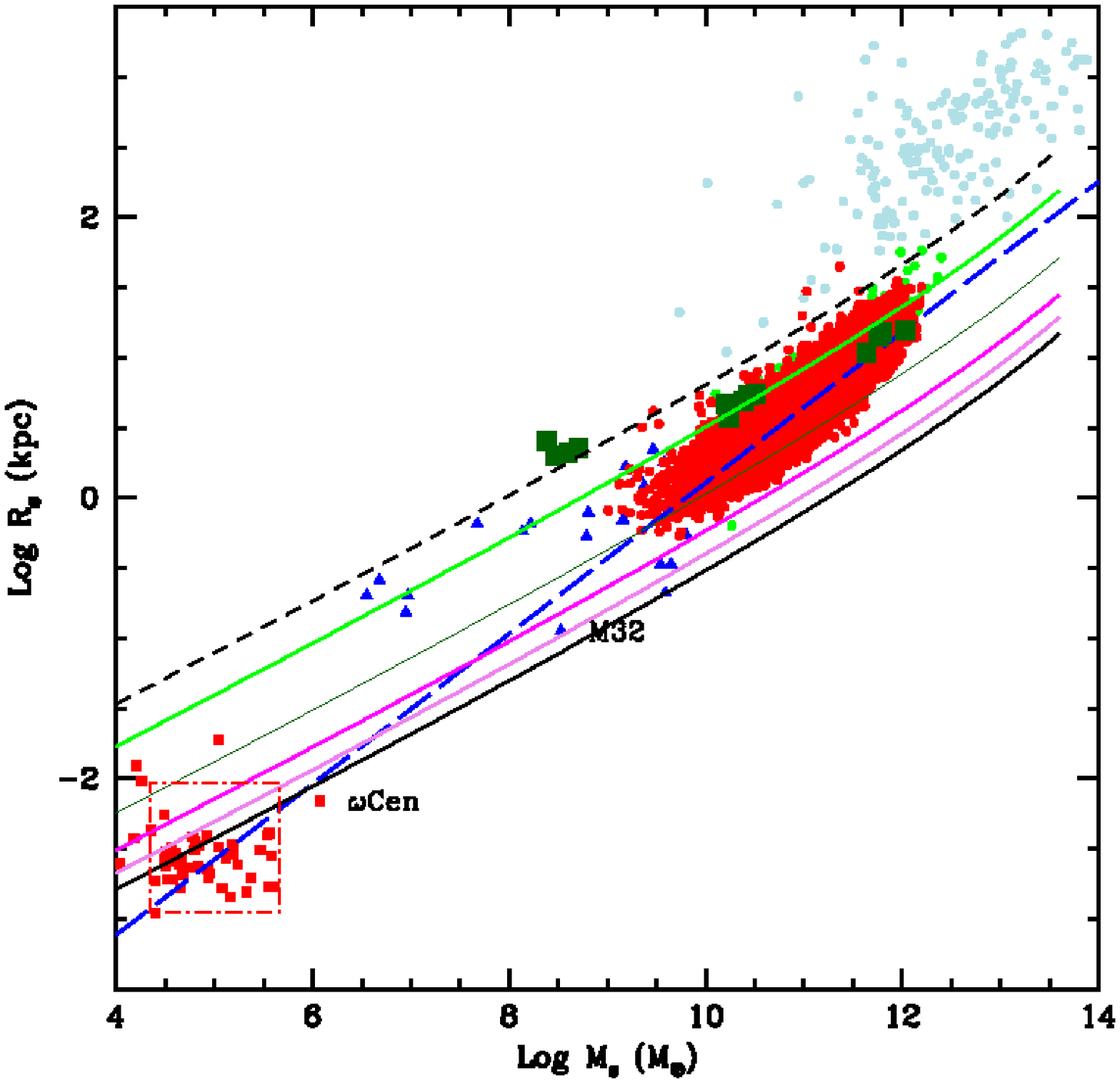}
\caption{ Comparison of the \citet{Fan_etal_2010} lines and the theoretical models by \citet{Merlin2012} 
(green filled squares)
with the observational data of \citet{Burstein_etal_1997} from GCs (small red squares), to DGs (small blue triangles), 
and GCGs (light blue 
filled circles) plus the ETGs of \citet{Bernardi_etal_2010} (red dots). The short-dashed blue 
line is the \citet{Fan_etal_2010}  relation for $z = 0$. The solid thin lines of different colors are the 
same 
relation but 
 for $z=1$, $z=5$, $z=10$,
$z=15$, and $z=20$, from  top to bottom in the order. The  long-dashed, thick blue line is the best-fit of the ETGs of 
\citet{Bernardi_etal_2010} extended to GCs and GCGs.}
\label{NBTSPH_models}}
\end{figure}

\section{Theoretical predictions for the MRR }\label{Theo_MRR}

In this section, we examine the theoretical foundations of the MRR and highlight the possible physical
causes of its occurrence.

\subsection{The  MRR of collapsing proto-galaxies made of DM + BM}\label{Collapsing_ProtoGal}

Independently of the formation scheme (either monolithic or hierarchical) the seeds of galaxy  structures are
perturbations of matter
made of DM and BM   that  undergo collapse when the density contrast with respect to
the surrounding medium  reaches a suitable value. Assuming spherical symmetry for the sake of simplicity,
the  MRR  for individual galaxies is given by

\begin{equation}
\left( \frac{4\pi}{3} \right) R_{D}^3 = \frac{M_{D}}{ \lambda  \rho_u(z)}
\label{mr2}
\end{equation}

\noindent where $\rho_u(z) \propto (1+z)^3$
is the density of the Universe at the redshift $z$, and $\lambda$  the factor for  the density contrast of
the DM halo. This expression is of general validity whereas the  function $\lambda$  depends on the
cosmological model of the Universe, including the $\Lambda$-CDM case. All details and demonstration of it can
be found in \citep[][ their Eq. 6]{Bryan1998}.

In the context of the $\Lambda$-CDM cosmology, \citet{Fan_etal_2010} have adapted the general relation
(\ref{mr2})
to provide an expression  correlating the halo mass $M_{D}$ and the  star mass $M_s$ of the galaxy born
inside it,  the half light (mass) radius $R_{s}$ of the stellar component, the redshift at which the
collapse takes place $z_f$, the shape of the BM galaxy via a coefficient $S_S(n_S)$ related to the Sersic
brightness profile from which the half-light radius is inferred and the Sersic index  $n_S$,  the velocity
dispersion of the BM component with respect to that of DM (expressed by the parameter $f_\sigma$), and
finally the ratio  $m=M_{D}/M_s$. The expression  is

\begin{equation}
R_{s}=0.9 \left(\frac{S_S(n_S)}{0.34} \right) \left(\frac{25}{m}\right) \left( \frac{1.5}{f_\sigma} \right)^2 
\left( \frac{M_{D}}{10^{12}  M_\odot} \right)^{1/3} \frac{4}{(1+z_{f})}.
\label{mr3}
\end{equation}

\noindent Typical value for the coefficient $S_S(n_S)$ is 0.34.
Furthermore, $f_\sigma$ yields the three dimensional star velocity dispersion as a function of the DM velocity
dispersion, $\sigma_s=f_\sigma \sigma_{DM}$. Here we adopt $f_{\sigma}=1$.  For more details see
\citet{Fan_etal_2010} and references therein.

The most important parameter of eq.(\ref{mr3}) is the ratio $m= M_{D}/M_s$. 
Basing on the Illustris data we 
have investigated how this ratio varies in the mass interval  $10^{8.5} <  M_{D} <10^{13.5}$ (masses are in 
$M_\odot$) and from $z=0$ to $z=4$ and made some predictions over a wider mass range $10^{4} <  M_{D} 
<10^{15}$ and a wider redshift interval from $z=0$ to $z=10$. The analysis and results are presented in 
Appendix \ref{Appendix_B} where we confine $m(M_D, z)$ within a rather narrow range of possible values and
suggest that a simple relationship giving the ratio $m$ as a function of $M_{DM}$ might be the one of 
eq.(\ref{final_m}) that is repeated here for the sake of clarity 

$$ \log m = \log \frac{M_D}{Ms}  =  0.062\, \log M_D + 0.429 $$
which turns out to be sufficiently accurate for a qualitative nature of our investigation.
In a very recent study \citet{Girelli_etal_2020} have thoroughly investigated the  stellar-to-halo mass 
ratio of galaxies ($M_s/M_{DM} = 1/m$ i.e. the inverse of our parameter $m$) in the mass interval $10^{11} < 
M_{DM} < 10^{15} $ and  redshifts from $z=0$  to $z=4$. 
They have used a statistical approach to link the observed galaxy stellar mass function on the COSMOS field 
to the halo mass function from the $\Lambda$CDM-Dustgrain simulations and an empirical model to describe the 
variation of the stellar-to-halo mass ratio a a function of the redshift. Finally they provide  analytical 
expressions for the function $m(M_D, z)$. Noting that in the mass interval in common their  results are in 
substantial agreement with those inferred from the Illustris data,  we prefer to provisionally adopt here  
eq. (\ref{final_m}) for the sake of internal consistency with the theoretical galaxy models in usage.     

We would like to point out that  relation \ref{mr3} is strictly valid only for monolithic infall of BM into 
collapsing DM
potential wells.  Nevertheless, this formula provides a general reference to obtain the typical
dimension of a galactic system as a function of its mass and formation redshift. While adjustments are
possible, the general trend is well defined. However, some deviations from this law are possible and
expected,
e.g. for low redshifts. See below for further discussion.   The MR-plane of the hydrodynamical  models and the
loci expected for different redshift from eqn. (\ref{mr3}) above are shown in Fig. \ref{NBTSPH_models}
together with the ETGs by \citet{Bernardi_etal_2010}.

The slope of relation (\ref{mr3}) is nearly identical to  the one estimated from
the hydrodynamical  models; the small difference can be fully ascribed to the complex baryon physics, which causes the
stellar system to be slightly offset with respect to the locus analytically predicted from DM halos.
Therefore, a  model slope (close to 1/3)  different from that of the observational MRR is not the result of
inaccurate description of the physical processes taking place in a galaxy; on the contrary, it mirrors the
fundamental relationship between mass and radius in any system of given mean density. Indeed it is remarkable
that quite complicated numerical calculations clearly display this fundamental feature.
\textit{If this is the
case, why do real galaxies gather along a line with a different slope?}

What is still missing in the above
MRR is that galaxies (globular clusters and cluster of galaxies) form and evolve in a given cosmological
scenario which ultimately drives the demography of objects over a large range of mass and dimensions in any
given volume of arbitrary size of the Universe. In other words, the real MRR is given
by the convolution of the MRR of each component with the underlying cosmological scenario that determines the
the mass interval spanned by galaxies at each redshift and the the relative percentage of galaxies of a
certain mass with respect to the others (otherwise known as halo mass function).

\begin{table*}
\begin{center}
\caption{ Coefficients of the polynomial interpolation of the relation (\ref{Lukic_interp}),
which provides the number density of haloes $n(M_{DM}, z)$ per (Mpc/h)$^3$.}
\begin{tabular}{|c|r|r|r|r|r|}
\hline
Mass $[M_\odot/h]$  &     A$_4$      &    A$_3$       &    A$_2$       &    A$_1$    &    A$_0$ \\
\hline
 5e7  &-2.34275e-5  & 	1.28686e-3   &	-2.97961e-2   &   2.11295e-1   &  2.02908  \\
 5e8  &-2.76999e-5  & 	1.49291e-3   &	-3.47013e-2   &   2.13274e-1   &  1.13553  \\
 5e9  &-1.31118e-5  & 	6.50876e-4   &	-2.36972e-2   &   1.31993e-1   &  0.23807  \\
 5e10 &-1.18729e-5  & 	6.65488e-4   &	-3.17079e-2   &   1.30360e-1   & -0.59744  \\
 5e11 &-1.47246e-5  & 	8.10097e-4   &	-4.65279e-2   &   1.13790e-1   & -1.44571  \\
 5e12 & 6.59657e-5  &  -7.19134e-4   &  -6.99445e-2   &   1.06782e-1   & -2.45684  \\
 5e13 &-7.34568e-4  &   9.99022e-3   &  -1.65888e-1   &  -9.48292e-2   & -3.11701  \\
 5e14 & 4.89975e-3  &  -5.17004e-2   &  -1.61508e-1   &  -5.83065e-1   & -4.28270  \\
\hline
\end{tabular}
\end{center}
\label{coef_lukic}
\end{table*}

\subsection{ The MRR from DM Halo Growth Function $n(M_{D},z)$}\label{MRR_DM_HGF}

The similarity of the MRR passing from star clusters to single galaxies of different mass and
morphological type and eventually to galaxy clusters  suggests that a deep relation
exists between the  way  all these objects populate the MR-plane and the
cosmological growth of DM halos.

\begin{figure}
\centering{
\includegraphics[width=8.0cm,height=8.0cm]{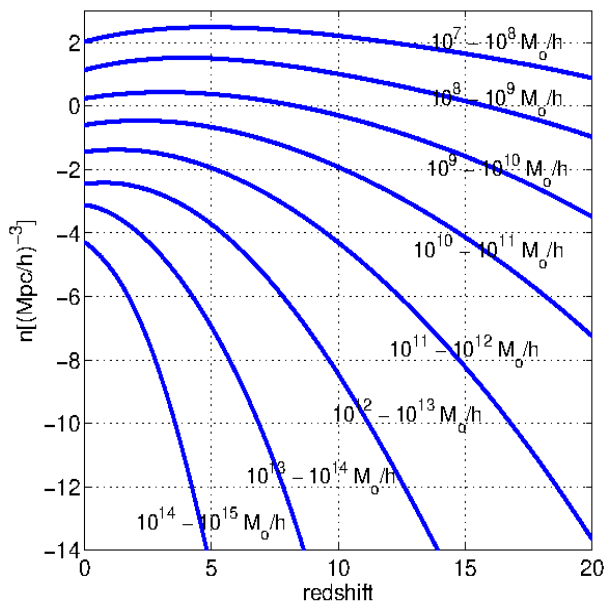}
\caption{The growth function of halos $n(M_{D}, z)$ reproduced from \citet{Lukic2007}. }
\label{lukic}}
\end{figure}

The distribution of the DM halo masses and their relative number density as a function of the redshift
has been the target of numberless studies which culminated with large scale simulations of the
structure of the Universe,
we quote here one for all, i.e. the Millennium Simulation \citep{Springel_etal_2005}. In parallel many
 studies have investigated
the so-called \textit{halo growth function, HGF}  as the integral of the \textit{halo mass function, HMF}.
Among others \citep[see for instance][]{Angulo_etal_2012, Behroozi_etal_2013} we recall
and make use of the results by \citet{Lukic2007} who, using  the $\Lambda$-CDM cosmological
scenario and the HMF of \citet[][]{Warren2006}, derive the HGF
$n(M_{DM}, z)$. This gives the number density of halos of different masses  per (Mpc/$h)^3$ resulting by all
creation/destruction events. The growth function is expressed  in terms of the normalized Hubble constant
$h=H_0/100$, where $H_0$ is assumed to be
$H_0$=70.1 Km/s/Mpc. The explored interval of redshift goes from 0 to 20.
The  $n(M_{DM}, z)$ function of \citet{Lukic2007} is shown in Fig. \ref{lukic}\footnote{This has been derived
from an analytical interpolation of the  data presented in Fig. 1 of \citet{Lukic2007}.}. Similar HGF are by
\citet{Angulo_etal_2012} \citet{Behroozi_etal_2013}. For more details, see \citet{Chiosi_etal_2012}.

The analytical representation of the $n(M_{D}, z)$ function displayed in Fig. \ref{lukic} is given by

\begin{equation}
n(M_{D}, z) = \sum_{n=0}^4 A_n(M_{D}) \times z^n
\label{Lukic_interp}.
\end{equation}

\noindent
where the  coefficients $A_n(M_{D})$ are listed in Table \ref{coef_lukic}.
Then we  count the total
number of halos per mass-bin $\Delta \log M_{D}$ at redshift $z=0$. This is simply given   by reading off
the values of the curves along the $y$-axis and interpolating for intermediate values. These are the  halos
that would nowadays populate  the synthetic
MR-plane and that should be  compared with the observed galaxies.
Since  the total number of halos read off the \citet{Lukic2007} plot refers to a volume of 1 (Mpc/$h)^3$,
any meaningful  comparison with observational data requires a suitable scaling of the theoretical values
by a suitable factor to match the real volume of the observed portion of the sky from
which the data are obtained (see below).

Although what we are going to say is well known, see the pioneer study of \citet{Press1974}
and \citet[][ for ample referencing]{Lukic2007}, for the sake of clarity and as relevant to our discussion we
note the following: (i) for each halo mass (or mass interval) the number density is small at high redshift,
increases to high values toward the present, and depending on the halo mass either gets a maximum value at a
certain redshift followed by a decrease (typical of low mass halos) or it keeps increasing as in the case of
high mass halos. In other words, first creation of halos of a given mass (by spontaneous growth of
perturbation to the collapse regime) overwhelms their destruction (by mergers), whereas the
opposite occurs for low mass halos past a certain value of the redshift; (ii) at any redshift high mass
halos are orders of magnitude less frequent than the low mass ones; (iii) at any redshift, the mass
distribution of halos has a typical interval of existence whose upper mass end (cut-off mass) increases at
decreasing redshift.

Given a certain number density of halos $N_s$, on the
$n(M_{D}, z)-z$ plane of Fig. \ref{lukic} this would correspond to an horizontal line intersecting the
curves
for the various  masses at different redshifts, i.e. obeying the equation $n(M_{DM}, z)= N_s$. Each
intersection provides a pair ($M_{DM}$, $z$) which gives the mass of the halos fulfilling the condition
$N_s=n(M_{D}, z)$ at the  corresponding redshift $z$ (or vice-versa the redshift satisfying the
condition for each halo
mass). For any value $N_s$ we get an array of pairs ($M_{D}$, $z$) that can be extrapolated to a continuous
function that,  with the aid of the \citet{Fan_etal_2010} relationship (in which the parameters $m$ and
$f_\sigma$ are fixed), provides the  corresponding relationship between the mass in stars and the half-mass
radius  of the baryonic galaxy associated to a generic host halo to be plotted on the MR-plane.

Repeating the procedure for different values of $N_s$, we get a manifold of curves on the MR-plane. It turns
out that with the $N_s$ corresponding to $10^{-2}$ halos per (Mpc$/h)^3$ (that roughly corresponds to the volume 
surveyed by the SDSS), the curve is just at the
edge of the observed distribution of ETGs on the  MR-plane. Higher values of $N_s$ would
shift it to larger halos (baryonic
galaxies), the opposite for lower values of $N_s$.  Why is $N_s=10^{-2}$ halos per (Mpc$/h)^3$ so
special? Basing on crude,  simple-minded arguments we recall that the total number of galaxies observed by
the
SDSS amounts to about $\simeq 10^6$, whereas the volume of Universe covered by it is about $\simeq 1/4$ of
the
whole sky times a depth of $\simeq 1.5\times 10^9$ light years, i.e. $\simeq 10^8$ Mpc$^3$, to which the
number density of about $10^{-2}$ halos per (Mpc$/h)^3$ would correspond\footnote{We are well aware that this
 is a very crude estimate not taking into account many selection effects both in the observations and in the
halo
statistics based on NB simulations, such as  the \citet{Lukic2007} plane itself. However, just for the sake
of
argument, we can consider it  as a good estimate to start with.}.

The equation $n(M_{DM}, z)= N_s$ with $N_s =10^{-2}$  or equivalently $10^6$ halos per $10^8$ Mpc$^3$
recast to  derive the halo mass $M_{DM}$ as a function of $z$ becomes
\begin{equation}
 \log M_{D} = 0.0031546\, z^3 - 0.006455\, z^2  - 0.183\, z + 13.287
 \label{eq_ns}
\end{equation}

\noindent
from which we get $M_D$. With the aid of eq.(\ref{final_m})  we derive the quantity $m$ and from its 
definition we 
obtain $M_s = M_{D}/ m$.  Finally, from  eq. (\ref{mr3}) we derive the associated radius  $R_s$.

The analytical fit of the MR relation determined in this way  and limited to the mass interval
$9.5 \leq \log M_s \leq 12.5$
($M_s$ in solar units) is

\begin{flalign}
   \log R_{s} =&  0.048562 (\log M_s)^3 -1.4329 (\log M_s)^2 \nonumber \\
                 &+ 14.544 (\log M_s) -50.898.
                 \label{shepherd_rel}
 \end{flalign}

\noindent This relation is meant to fit the distribution of the sole ETGs on the MR-plane. We note that
the slope gradually changes from 0.5 to 1 and above as we move from the low mass to
the  high mass range. It is worth recalling here that a similar trend for the slope is also indicated
by the
observational data \citep[see][and references therein]{vanDokkum2010}. Owing to the many uncertainties we do
not try to formally fit the median of the empirical MRR, but we limit ourselves to show that the locus
predicted by $N_s= 10^{-2}$ halos per (Mpc$/h)^3$  falls on the MR-plane close to the
observational MRR. Lower or higher values of the halo number density would predict loci in the MR-plane too
far from the observational MRR.

\begin{figure}
\centering{
\includegraphics[width=8.0cm,height=8.0cm]{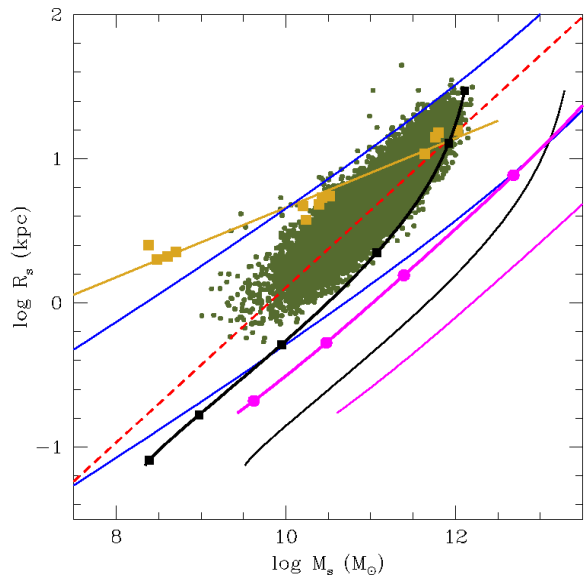} }
\caption{The \textit{Cosmic Galaxy Shepherd} (CGS) and the corresponding locus of DM parent halos(the black
solid thick and thin lines, respectively) for the number density of  $N_s = 10^{-2}$ halos per (Mpc/h)$^3$. In
addition to this we show the case with $N_s = 10^{-8}$ halos per (Mpc/h)$^3$ (the magenta solid  thick and 
thin  lines). The various loci are plotted onto the observational MR-plane (see text for all details). We
also draw the observational data for the  HB sample of \citet{Bernardi_etal_2010} with their linear fit,  
and two theoretical MRR from eqn. (\ref{mr3}) by \citet{Fan_etal_2010} (solid thin blue lines),
relative to $z=0$ and $z=10$, with $m=15$ and $f_\sigma=1$. In addition to this, the present day position of 
the  reference galaxy models and their linear fit are shown (the golden filled squares and  solid line). 
Finally, the filled black squares and magenta circles show the CGS at different values of the redshift (z=0, 
1, 5 10, 15, and 20) for the two cases of the number density per 
(Mpc/h)$^3$. }
\label{reffmass1}
\end{figure}

Finally, we call attention on the fact that the locus on the MR-plane defined by the relation (\ref{shepherd_rel})
is ultimately related to the top end of the mass scale of halos (and their associated baryonic objects) that
can exist at each redshift. In other words, recalling that the mass of any intersection pair for
$N_s=10^{-2}$
corresponds to halos becoming statistically significant in number on the observed spatial scale at the
associated redshift, this can be interpreted as the so-called cut-off mass in the \citet{Press1974} or
equivalent formalisms \citep[see][ for details and references]{Lukic2007}.
Therefore, this provides also an upper boundary to the mass of galaxies that are allowed to be in place (to
collapse) at each redshift. We name these locus the {\it Cosmic Galaxy Shepherd} (hereafter CGS). All this is
shown in Fig. \ref{reffmass1}, where we also plot the curves relative to $N_s= 10^{-8}$ halos per
(Mpc$/h)^3$, corresponding to 1 halo per $10^8$ (Mpc$/h)^3$, for the sake of
comparison.

There are two points to be clarified. First, this way of proceeding implies that \textit{each halo hosts one
and only one galaxy and that this galaxy is an early type object matching the selection criteria of the
\citet{Bernardi_etal_2010} sample}. In reality ETGs are often seen in clusters and/or groups of galaxies and many
large spirals are present. Only a fraction of the total population are  ETGs. One could try to correct for
this issue by introducing some empirical statistics about the percentage of ETGs among all types of galaxy.
Despite these considerations, to keep the problem simple we ignore all this and  stand on the minimal
assumption that each DM halo hosts at least one baryonic component made of stars. This is a strong
assumption,
on which we will come back again later on. Second, we have assumed that  $m$ varies with the halo mass. According to
\citet{Fan_etal_2010}, the empirical estimate of $M_{DM}/M_s$ ration is about 20-40, our estimate yield a
mean value $\langle m \rangle \simeq 20$. However, a slightly higher value for
$m$ does not invalidate our analysis, because it would simply shift the location of the baryonic component on
the MR-plane corresponding to a given value of $N_s$. Finally, $f_\sigma=1$ is a conservative choice.
The same considerations made for $m$ apply also to this parameter.

Along the line for the Cosmic Galaxy Shepherd, redshift and cut-off mass go in inverse order, i.e. low masses
(and hence small radii) at high
redshift and vice-versa. This means that a manifold of MRRs defined by eqn. (\ref{mr3}), each of which
referring to a different collapse redshift, can be  selected, and along each MRR only masses (both parent
$M_{DM}$ and daughter $M_s$) smaller than the top end are permitted, however each of which with a different
occurrence probability: low mass halos are always more common than the high mass ones. In the observational
data,  it looks as if ETGs should occur only towards the high mass end of each MRR, i.e. along the locus on
the MR-plane whose right hand side is limited by the CGS. This could be the result of  selection effects,
i.e.
(i) galaxies appear as ETGs only in a certain interval of mass and dimension and outside this interval they
appear as objects of different type (spirals, irregulars, dwarfs etc..), or (ii) they cannot even form or be
detected (e.g. very extended objects of moderate/low mass). Finally, in addition to this, we  argue that
another physical reason limits the domain of galaxy occurrence also on the side of the low mass, small
dimension objects. We will come to this later on.

\subsection{More on the cosmic galaxy shepherd}\label{cosmic_shepherd}

If we compare the present-day position of the reference hydrodynamical models on the MR-plane with the region
populated by real galaxies (see Fig. \ref{composite_RsMs} and/or Fig. \ref{NBTSPH_models}), at a first glance 
one would be tempted to conclude that only the high mass models suited to massive ETGs fairly agree with 
observations, whereas the low mass ones (and to some extent also those of
intermediate mass) apparently have too large radii with respect to their masses. However, before drawing the
conclusion that essentially the models fail to reproduce the data, it is worth recalling that a the
observational distribution of galaxies on the MR-plane result from the combined action of may factors not
yet taken into account by our analysis. Is the observational sequence of ETGs populated only by galaxies 
behaving as our massive ones? Or what else? Since for any value of the halo mass there is a certain redshift 
below which halos of this mass start decreasing in number by mergers \citep{Lukic2007}, galaxies generated by 
those halos become more and more unlikely as it should be the case for our low mass models. Indeed when at a 
given   redshift we have assumed the existence of halos of any mass, we have neglected this important effect. 
Therefore, the situation may occur that halos/baryonic model galaxies are calculated and plotted
onto the MR-plane even though according to the above arguments their existence is very unlikely.

On this ground, we argue that the observational MRR of ETGs (galaxies in general) is the result of
convolving two agents: the halo growth function providing the number density of halos of different mass  as
a function of the redshift (in the concordance $\Lambda$-CDM Universe), and the fundamental MRR determining
the probable size of a galaxy as a function of its mass and formation redshift after the collapse.

\section{Tightening things up in cosmological context}\label{MRR_cosmo}

\begin{figure*}
\centerline{
\includegraphics[width=12cm,height=12cm]{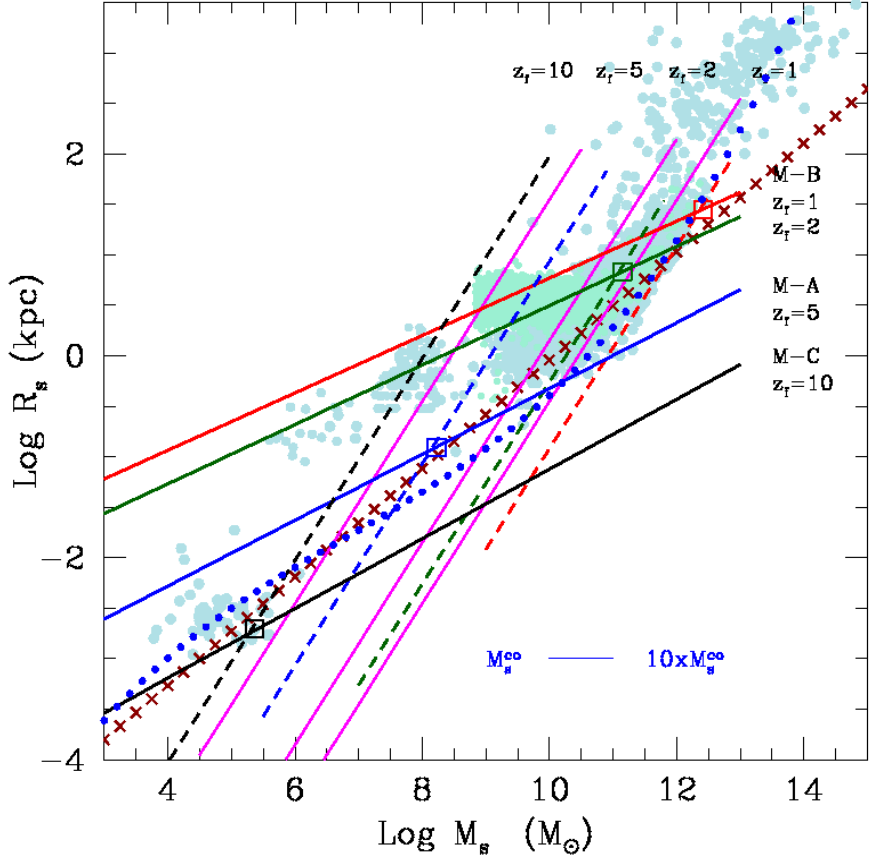}  }
\caption{ The Mass-Radius Relationship: comparison between data and theory.  Radii $R_{s}$ and stellar masses $M_s$ are in kpc and  
$ M_{\odot}$, respectively. The pale-blue filled circles are all
the data considered in this study,  the pale green filled circles  the models of Illustris. 
The dark-red thick  crossed line is the linear best fit
of  normal ETGs ($M_T \geq 10^{10}\, M_{\odot}$) given by eq.(\ref{RsMs_Bernardi}), however
prolonged down to the
region of globular clusters and upwards to that of galaxy clusters.
The four long-short dashed lines labeled Mod-A ($z_f =5$, blue), Mod-B ($z_f=1$, red and $z_f=2$, dark green) and Mod-C 
($z_f=10$ black) are the analytical
 relationships of eq. (\ref{re_mstar}) showing the loci of galaxy models with different mass but constant initial density 
 for different values of
redshift of galaxy formation $z_f$ as indicated. These lines are the best fit of the the models by 
\citet{Chiosi_Carraro_2002}, \citet{Merlin2012} 
and \citet{Chiosi_etal_2012}, see Tables \ref{tab0}, \ref{tab1}, and
\ref{tab2}.
The magenta solid lines are the MRRs of
galaxies in virial conditions and with different velocity dispersion (50, 250 500 km/s from left to right)
The dashed black lines labeled by different
values of $z_f$ are the MRRs expected for galaxies with total mass equal to $50\times M^{CO}(z)$,
the cut-off mass of the Press-Schechter at varying $z_f$ according to relation (\ref{m_knee}). 
 These lines go in pairs with the best-fit lines of models of identical initial 
density labeled by the same color.  
The large empty squares visualize the intersections between the lines of constant initial  density and the MRRs for
$50\times M^{co}$ galaxies for equal values of the redshift. All the intersections  lie very close to the
relation of eq.(\ref{RsMs_Bernardi}) shown by the dark-red crossed line. This is the linear interpretation of 
the observed MRR, i.e. the analytical demonstration of Section \ref{MRR_cosmo}.
Finally, the curved blue dotted  line  shows the expect MRR from the cosmological distribution of
DM halos and stellar contents in turn of different mass by \citet{Lukic2007}, however extended to the domain of globular clusters 
and galaxy clusters (see the text for more details). 
Note the ever changing slope
decreasing with the halo mass and the stellar mass turn in. Remarkably the curved line first runs very close to the
large empty squares, second to linear fit of the data (crossed line), and third  accounts for the observed MRR passing 
from globular clusters to galaxy clusters (about
ten orders of magnitude difference in the stellar mass).  Finally, the horizontal blue
line shows the interval for $M_s$ corresponding
to  initial masses $M^{CO}(z) < M_T < 10\times M^{CO}(z)$ (the percentage of galaxies in this interval amounts to $\simeq 15\%$) 
highlights that at each redshift the high-mass  edge of the MRR is not a 
sharp border.
}
\label{mass_radius}
\end{figure*}

In this section we seek a common explanation for the observational distribution of astronomical objects
going from GCs, to galaxies like DGs, ETGs (and to a less extent  also Spiral Galaxies) to
Clusters of Galaxies, the mass of which  spans about eleven orders of magnitude. The situation is
shown in  Fig. \ref{mass_radius}. The pale-blue filled circles are the
normal/giant ETGs,    the DGs,   the  GCs, and the GCGs. No distinction is made among different groups 
or sources of data. The aim here is to qualitatively display the region of the MR-plane populated by real objects 
of different mass, size and morphological type.

Let us quickly summarize once more the  main features of the distribution:

(i) The family of GCls is well detached from the body of normal/giant ETGs (let us say those
with mass larger than about $ 10^{10}\,\, M_{\odot}$). However,  the region in between is populated by DGs.
At the top of the distribution there are the GCGs  with the largest radii and masses. The richest
sample to our disposal is made of ETGs (the Spiral Galaxies occupy more or less the same region). The
relative number of objects per group is not indicative of the real number frequencies because severe
selection effects are present. The best fit of the ETGs data from the various sources yields the relations
of  eqn.(\ref{RsMs_Burstein}) for \citet{Burstein_etal_1997},   eqn.(\ref{RsMs_Bernardi}) for \citet{Bernardi_etal_2010}
and eqn.(\ref{mr_wings1}) for the WINGS data, in which only objects with $M_s \geq 10^{10}\, M_\odot$ are considered not to contaminate
the samples with DGs. Since the  slopes differ by 0.1 and the zero-points  by 0.88, we consider the three 
relationships fully equivalent.

(ii) If we extrapolate any of the relation above holding for massive ETGs downward to the mass range of
GCs and upward to that of GCGs, we see that the same relation   provides a
lower limit
to GCls,  passes through  $\omega$Cen  and M32, provides the lowest limit to the
distribution of DGs,  and finally  hit the region of GCGs.

(iii) There are  no galaxies in the semi-plane for  radii $R_s$ smaller than the values fixed by relation
(\ref{RsMs_Bernardi}), independently of the galaxy mass, but for the so-called ``compact galaxies'' that
we will examine in a forthcoming paper (Chiosi et al. 2020 in preparation).

(iv) Starting from the cosmological  HMF we have been able to derive a MRR, named Cosmic Galaxy Shepherd,
providing a sort of
mass limit in the MR-plane to the distribution of ETGs. The analytical expression for this limit is given
by eq.(\ref{shepherd_rel}) and it plays the same role as the three MRR above. The only difference is
that it gradually changes  its slope from $\simeq$0.5 to  $\simeq 1$ at increasing the galaxy
mass. Extending the Cosmic Galaxy Shepherd down to GCs and up to GCGs  a different
analytical approximation is possible

\begin{flalign}
 \log R_s =&0.007584 (\log M_s)^3 - 0.1874\,(\log M_s)^2  \nonumber \\
                   & + 1.908 (\log M_s) - 9.027
\label{good_relation}
\end{flalign}
with $R_s$ and $M_s$ in the usual units. This line is the
analog of the linear global fit above. As already said it represents the cut-off mass of the halo
distribution function at varying  redshift however translated onto the $R_s$ vs $M_s$ plane. 
This coincidence provides a profound physical meaning to the
transverse line splitting the MR-plane in two regions, i,e. the region in which the vast majority of
galaxies are found and the region of avoidance.

(v) Galaxy models tell a more complicated situation. The monolithic
hydrodynamical models
by \citet[][]{Chiosi_Carraro_2002} and the series of early-hierarchical models by
\citet[][shortly indicated Mod-M]{Merlin2012} yield the following MRRs

\begin{eqnarray}
      \log R_s& =&   0.331 \log M_{s}  - 3.644       \qquad {\rm \,\, Mod-A  } \label{mr_3a}  \\
      \log R_s& =&   0.273 \log M_{s}  - 1.994       \qquad {\rm \,\, Mod-B}   \label{mr_3b} \\
      \log R_s& =&   0.241 \log M_{s}  - 1.750       \qquad {\rm \,\, Mod-M }  \label{mr_4}
\end{eqnarray}

\noindent
It is worth noting that  both the slope and zero point  of models Mod-A and Mod-B change with the redshift,
models Mod-M are similar to models Mod-B, and finally  the variation in slope is smaller than that in zero-point.
Recalling that the three groups of models are calculated with different redshift of galaxy formation (hence 
initial density)
but similar internal physical processes, this means that the slope is fixed by the physical structure of the 
models,  whereas the
zero-point is reminiscent of the initial density.  The  slope of the above relations is not identical to
that of ETGs, relation  (\ref{RsMs_Bernardi}), but more similar to that of DGs.
However, along the sequence of each group, the most massive models in which star formation is terminated  fall into the
region of ETGs. 

The Illustris models yield similar relationships, once they are split in two groups
\begin{equation}
      \log R_s =   0.297 \log M_{s}  - 2.513       \qquad {\rm for \,\, \log M_s \leq 10.5 }
      \label{mr_5b}
\end{equation}
and \begin{equation}
      \log R_s =   0.519 \log M_{s}  - 4.492       \qquad {\rm for \,\, \log M_s \geq 10.5 }
      \label{mr_5a}
\end{equation}
The first relation holds for the vast majority of models and reminds the one of normal DGs, whereas the second
relation holds for a small group of objects and is close to the case of ETGs. Furthermore, the models of the first
group with the MRR of eq. (\ref{mr_5b}) are the seeds of bigger galaxies, which after reaching a suitable
value by mergers  and terminating all star formation activity, give origin to galaxies located along the
MRR of eq (\ref{mr_5a}).

Finally, the  MRR of eq.(\ref{mr3}) of \citet{Fan_etal_2010} with slope 0.333 is
nearly identical to that of theoretical models, i.e.  eqns. (\ref{mr_3a}), (\ref{mr_3b}), (\ref{mr_4}), and
(\ref{mr_5b}). In other words, by construction the \citet{Fan_etal_2010} lines visualize galaxies born at the
same redshift but with different masses. The most important issue here is  ``Why observational MRRs for ETGs
are so different from the theoretical ones but for the most massive objects?''

(vi) To answer the above question, we start from the following general considerations. It goes without
saying that the gravitational collapse of proto-clouds giving origin to a galaxy
is in general accompanied by important side phenomena such as star formation and consequent energy feed-back,
gas cooling and heating,
galactic winds removing energy and mass, mass and energy acquisition by mergers, etc. Therefore the
theoretical models may change
depending on the detailed physical description of all these energy producing/removing phenomena together with
those for
the mass acquisition/loss. In this scenario, the ideal reference galaxy formation model would be the
dissipation-less collapse of DM+BM halos originated from  primordial
density perturbations of rms amplitude toward the equilibrium structure \citep{Gott_Rees1975,
Faber1984,Burstein_etal_1997}. In brief, if $\delta$ is the rms amplitude of primordial density
perturbations of total mass $M_T=M_{DM} + M_{BM}$
\begin{equation}
      \delta \propto   M_T^{- \frac{1}{2} - \frac{n}{6} }
\label{delta_M}
\end{equation}
\noindent where $M_T$ is the mass at the initial red-shift, and $n$ is the slope of the density fluctuation
$\delta$. After collapse, the equilibrium structure of a halo originated from given $\delta$ and $M_T$
follows the relations \citep{Gott_Rees1975}

\begin{equation}
      R_T   \propto   \delta^{-1} M_T^{ \frac{1}{3} }
\label{delta_MR1}
\end{equation}
\noindent from which we immediately get

\begin{equation}
      R_T   \propto M_{T}^{ \frac{5+n}{6}}
\label{delta_MR2}
\end{equation}
\noindent Inserting $n=-1.8$, the power spectrum of CDM
\citep{Blumenthal_etal_1984}, we get the  relation

\begin{equation}
      R_{T}   \propto M_{T}^{0.53} \
\label{delta_MR3}
\end{equation}
\noindent The slope of the MRR derived from the dissipation-less collapse is about same of
eqn. (\ref{RsMs_Bernardi}) for ETGs, whereas the proportionality constant cannot be fixed by these simple arguments. 
For the sake of a simple discussion, we approximate  $M_T\simeq M_D$ and $R_T \simeq R_D$ and replace eqn. \ref{delta_MR3} with
\begin{equation}
      R_{D}   \propto M_{D}^{0.53} \
\label{delta_MR4}
\end{equation}
Inside this halo up a galaxy made of stars is built  up over years with mass $M_s$ and half-mass radius $R_s$. Using the 
Illustris models (see Sect. \ref{models_illustris}) we  may derive the relationships   $\log R_s = \gamma \log R_D +\eta$ 
and $\log M_s = \alpha \log M_D + \beta$.  
Inserting these relations into eqn.(\ref{delta_MR4}) we obtain

\begin{equation}
\log R_s = 0.53 \frac{\gamma} {\alpha} \log M_s -0.53 \frac{\gamma \beta}{\alpha} - {\eta}  + k \gamma   
\label{delta_MR5}
\end{equation}
where the constant $k$ owes its origin to the initial indeterminacy of the proportionality factor in eqn. (\ref{delta_MR4}).
Limiting to models at $z=0$ and to the region of the MR-plane in which the MRR is evident
(roughly for $\log M_s \geq 10.5$)  we get  $\alpha=0.533$,  $\beta=4.607$,  $\gamma = 0.544$, and $\eta = -0.103$, whereas for
smaller masses ($\log M_s < 10.5$ we obtain $\alpha=0.883$,  $\beta=0.088$,  $\gamma = 0.171$, and $\eta = 0.377$. With these 
values we have 
\begin{eqnarray}
\log R_s &=& 0.541 \log M_s  -4.702 + k\gamma   \quad  {\rm for} \log M_s > 10.5   \label{delta_MR6a} \\
\log R_s &=& 0.102 \log M_s  -0.017 + k\gamma   \quad  {\rm for} \log M_s < 10.5   \label{delta_MR6b}
\end{eqnarray}
The  term $k \gamma$ cannot be determined unless the constant $k$ is specified by fixing the initial conditions of the 
collapsing proto-halo.
The slope of relation (\ref{delta_MR6a}) does not significantly differ from the one of eqn. (\ref{delta_MR3}) dissipation-less 
collapse, and eqn. (\ref{RsMs_Bernardi}), the empirical MRR of \citet{Bernardi_etal_2010}. This is possible only for 
the most massive galaxies of M-MRR manifold. For galaxies of smaller mass the final relation, eqn. (\ref{delta_MR6b}), 
largely departs from  it. 

Based on the above considerations, one might be tempted to conclude that equation (\ref{RsMs_Bernardi})
represents the locus in the MR-plane  of galaxies whose formation  closely follows the dissipation-less collapse 
from the initial state (mass, radius, mean density, etc.) to the final one indicated on the MR-plane by the 
final values of $M_s$ and $R_s$. Were this the case, special conditions ought to hold for all objects like the DGs that
clearly deviate from
this relationship. The explanation is different for  the monolithic and hierarchical scenario:

a) In the standard monolithic view, in addition to star formation and the various gas heating and
cooling processes, galactic winds are  another key ingredient to consider. The analysis made by
\citet{Chiosi_Carraro_2002} is particularly useful.  Following their reasoning, we derive the 
relationship between the gas mass lost in the wind as function of the mass in stars in the galaxy models, 
the low mas ones in particular because the DGs galaxies show the largest deviation from the observed MRR, 
eq. (\ref{RsMs_Bernardi}) or eq. (\ref{delta_MR3}). Using the entries of Table~\ref{tab0} in Appendix
\ref{Appendix_C} (those for Mod-B), we obtain the following relations

\begin{eqnarray}
\log (\frac{ R_{s}}{ R_{s}^o})  &=& - 0.264 \log M_s + 3.271   \\
\log (\frac{M_{g,w}}{M_{g,i}}) &=& - 4.038\log M_s + 5.490   \\
\log (\frac{ R_{s}}{ R_{s}^o})  &=&  1.065 \log (\frac{M_{g,w}}{M_{g,i}})  + 1.825
\end{eqnarray}

\noindent where $R_{s}$ is the current effective radius of the galaxy,  $R_{s}^o$ the
radius it would have if  strictly following eq. (\ref{RsMs_Bernardi}), and $M_{g,w}/M_{g,i}$
the gas mass lost in the galactic  wind normalized to the initial gas mass. The last relation shows   that
$\log  (R_{s}/ R_{s}^o)$ increases with $M_{g,w}/M_{g,i}$. The stronger the galactic wind, the larger is
the final effective radius. In other words, galaxies tend to depart from the locus represented by
eq. (\ref{RsMs_Bernardi}) and/or eq.(\ref{delta_MR3}   at increasing mass and galactic winds in turn, the low mass ones having 
the strongest effect. In addition to
this, we noted that the relative efficiency of galactic winds tends to decrease at
increasing initial density of the proto-galaxy(compare models A and B in  Table~\ref{tab0} of the Appendix \ref{Appendix_C}).
This means that the effect of galactic winds in inflating low mass galaxies of high initial density is lower
so that their final radius will be closer to the value predicted by eqs. (\ref{RsMs_Bernardi}) and/or
(\ref{delta_MR3}). {\it The conclusion is that the flatter slope of the theoretical MRR in the monolithic scheme is likely 
caused by  galactic winds}.

b) In the case of the hierarchical scenario, the situation is more complicated because mergers and galactic 
winds both concur to inflate  a galaxy.
To clarify the issue \citet{Chiosi_Carraro_2002} discussed the  case of a merger between two galaxies calculated  
by \citet{Buonomo2000} of which they  knew all details:  a $2\times 10^{11}\, M_{\odot}$  galaxy was simulated 
by merging two disc-like sub-units, each one with 
total mass (BM+DM) equal to $1\times 10^{11}\, M_{\odot}$, and ratio of BM to DM equal to 0.1.
At the time of encounter the mass in stars of each sub-unit was about $M_s=6\times 10^{8}\, M_{\odot}$ and the
mass in gas  $M_g=9.3\times 10^{9}\, M_{\odot}$. The merger was accompanied by little  star formation,
so that
the total mass in stars,  gas, and BM  of the newly formed galaxy was $M_s=1.2\times 10^{9}\, M_{\odot}$,
$M_g=1.8\times 10^{10}\, M_{\odot}$ and $2\times 10^{10}\, M_{\odot}$, respectively. The effective radius of the composite object was 
$R_{s}=8.4$ kpc and its   shape  reminded that of  an elliptical one. The
merger galaxy  was  shown in Fig. \ref{composite_RsMs} by the big empty star. It looks like our Mod-B of
comparable  total star mass   except for the fact that it is more diffuse. Basing on these calculations, we
estimate that  merging two single disc-like objects made of stars and gas to build up a galaxy with twice as
much total mass as the component galaxies would generate an elliptical-like galaxy, the  star mass
and effective radius of which are smaller and higher, respectively, by $\Delta M_s/M_s\simeq -0.9$ and $\Delta
R_{s}/R_{s}\simeq 0.5$ compared to the case of an elliptical of the same mass but obtained with  the
monolithic scheme. However, if  mergers might inflate the final object
\citep[see][]{Hernquist_1992,Hernquist_1993} and therefore  explain the position of DGs (see also the many  models 
of the Illustris database calculated in the hierarchical framework), they can hardly explain the observational MRR of ETGs. 
In other words, mergers may concur to build up a galaxy, but are not determining  the final MRR. 

On consideration of these premises, we suggest that the observational MRR of ETGs, either
eqn.(\ref{RsMs_Burstein}) or eqn.(\ref{RsMs_Bernardi})
or eqn.(\ref{mr_wings1}),  represents the locus on the
MR-plane  of galaxies whose present-day structure  closely mimics the ideal situation of mechanical (virial)
equilibrium and passive evolution. Their MRR is also the  boundary  between permitted and forbidden 
regions of the MR-plane. The DGs or less massive objects  have a different
interpretation, because they significantly depart from the above evolutionary scheme and the MRR holding for
ETGs.

(vii) To prove  the above statement we strictly follow the method proposed long ago by 
\citet{Chiosi_Carraro_2002} however updated to the recent theoretical and observational data. 
To this aim, we draw on  Fig.\ref{mass_radius} two loci and a mass interval as function of the
initial density (redshift):

(a) The first locus  is the present-day MRR traced by model galaxies of different mass but similar 
initial density at varying  redshift. Most of this issue has already been examined in detail but from a 
different perspective in Sect.\ref{compare_theor_obser}. The novelty here is to provide an easy to use 
relationship yielding the expected MRR of theoretical models as a function of the formation redshift $z_f$ 
for the bulk of the stellar content of a galaxy. Using all the models to our disposal, i.e. 
\citet{Chiosi_Carraro_2002} and \citet{Merlin2006,Merlin2007,Merlin2010,Merlin2012, Chiosi_etal_2012} 
we get the  relation
\begin{eqnarray}
\log R_{s} &=& [-1.172 - 0.412\, (1+z_f)] +  \nonumber \\
           & & [0.244 + 0.0145 \, (1+z_f)] \, \log M_{s}
\label{re_mstar}
\end{eqnarray}

\noindent
The following cases are shown in Fig. \ref{mass_radius}:  $z_f \simeq 1$,  $z_f \simeq
2$, $z \simeq 5$ and  $z_f \simeq 10$. The procedure is safe thanks to the
regular behavior of the models and the density-mass relationship of eq. (\ref{delta_MR3}).
This relationship  is compatible with the MRRs predicted by \citet{Fan_etal_2010}, see  eq.(\ref{mr3}).
Finally, they  also agree  with the theoretical models of Illustris by \citet[][]{Vogelsberger_2014a, 
Vogelsberger_2014b}.

(b) The second locus is the MRR for galaxies with the  statistical maximum mass allowed by the underlying 
HGM at any redshift. This relationship can be derived (either graphically or numerically) from any HGM 
in literature e.g. \citet{Lukic2007}, or \citet[][]{Angulo_etal_2012}, or \citet{Behroozi_etal_2013}.
In Sect. \ref{MRR_DM_HGF} we have derived this relation for the HGF of \citet{Lukic2007}: see the analytical 
expression of eq.(\ref{shepherd_rel} or its  extension to the whole MR-plane given by eq, 
(\ref{good_relation}).
However, for the sake of a very simple  yet  instructive analytical approach
we prefer  to make use here of the classical   \citet{Press1974} function of haloes as a sort
of mass distribution function  for galaxies:  in the simplistic assumption of one galaxy per halo  it provides the relative number of 
galaxies per mass bin at varying  redshift.  Furthermore, we  consider the cut-off mass
$M_T^{CO}$  of the \citet{Press1974} function as the maximum mass limit of the galaxy masses at the current  
value of the redshift.
According to the \citet{Press1974} formalism, varies  with the redshift according to

\begin{equation}
           M_{T}^{CO} = M_{N} \times (1+z)^{ - \frac{6}{ n + 3} }
\label{m_knee}
\end{equation}

\noindent
The exponent $n$  is the  slope of the initial power spectrum of the perturbations, $M_{N}$ is a suitable
normalization mass scale.  At any redshift, most of the galaxies have total masses smaller than $M_{T}^{CO}$, even
if higher values cannot be excluded. Indeed, the fractional  mass in (or fractional number of)  galaxies
with mass greater than $M_{T}^{CO}$ is a function of $n$. For $n$=-1.8, the percentage of galaxies in the mass
interval $M_{T}^{CO} < M_T < 10\times M_{T}^{CO}$ amounts to about 15\% and in the range $10\times M_{T}^{CO} < M_T <
100\times M_{T}^{CO}$ to about 1\%.  Therefore,  at any redshift galaxy masses up to say $10\times M_{T}^{CO}$
may occur with a sizable probability.

With the aid of the relationships presented in Tables~\ref{tab0}, \ref{tab1}, and \ref{tab2} of 
Appendix \ref{Appendix_C} (providing the
mass in stars and effective radius of BM as a function of $M_T$) and limited to the case with
$M_T = \gamma \,\, M_{T}^{CO}$ for $\gamma=10$  and   power spectrum $n=-1.8$, we get

\begin{equation}
  R_{s}= 16.9 \times 10^{12} \times \gamma^{-0.79}\times (1+z)^{3.96} \times M_{s}
\end{equation}

\noindent
where $R_{s}$ and $M_{s}$ are in kpc and $M_{\odot}$. These loci are shown in Fig. \ref{mass_radius} by the
dotted lines labeled by the redshift, namely $z\simeq 1$, $\simeq 2$, $\simeq 5$ and $\simeq 10$. {\it On the MR-plane, they
represent the rightmost extension of
the lines of constant density and maximum galaxy mass in turn. At decreasing redshift the
boundary progressively moves toward
higher masses}. Similar lines and conclusions can be recovered  using the HGF of \citet{Lukic2007} or 
\citet[][]{Angulo_etal_2012}, or \citet{Behroozi_etal_2013}.

\noindent
(c) Finally, the third locus is the expected interval for $M_{s}$ (the present day star mass in a galaxy) for
objects with total mass comprised between $M_{T}^{CO}$ and $10\times M_{T}^{CO}$ as a function of
the redshift. To this aim,  the relation $M_s(M_T)$ has been  plugged into relation (\ref{m_knee})  for
$M_{T}^{CO}$. The permitted intervals are  shown  in Fig. \ref{mass_radius}  by the horizontal lines labeled
$M_s^{CO}$. The interval for $M_s$ going from $10^{10}\,
M_{\odot}$ to $10^{12}\, M_{\odot}$  is fully compatible with a redshift interval for the formation of the  majority of stars in
a galaxy from 2
to 1. This interval is also the mass range over which at each redshift the probability for the occurrence of massive
galaxies falls to negligible values. In other words, on the MR-plane the right-hand border of the MRR has a 
natural width.

(viii) Having set the whole scene, we proceed to the final step. If our reasoning is correct the basic
relationship for ETGs, either eqs. (\ref{RsMs_Bernardi}) or (\ref{delta_MR3}),
however extended to the whole mass range of the objects under consideration (from GCs to GCGs) should
correspond to the intersection  of lines of constant initial density  and the lines $\gamma \, 
M_T=M_{T}^{CO}(z)$ for
equal values of the redshift, at least for all values of redshift greater than about 1. This indeed is what we
see in Fig. \ref{mass_radius} (the large empty squares). This result provides the analytical demonstration that the
observational MRR stems from the product of two main agents:  the mechanism governing galaxy formation and 
evolution (monolithic, early-hierarchical, or fully hierarchical) and  the HGF  and its variations with the redshift.
It follows from this analysis that the MRR is the locus of objects in mechanical (virial) equilibrium 
and passive or nearly passive evolutionary
stage. This MRR is mainly traced by ETGs, GCs, a few DGs and GCGs. The large majority of DGs lie above it.  
We have already argued about the physical reasons why DGs (but for  few exceptions) cannot yet  lie on this MRR. 
Ongoing active star formation and strong galactic winds. This finding confirms the result obtained by \citet{Donofrio_etal_2019b}: 
only passive galaxies (strongly decreasing their luminosity as the redshift goes to zero) trace the MRR  on the MR-plane with a 
slope 0.5 to 1, the highest value being reached by objects 
that suffered the  strongest luminosity decrease with the redshift, i.e. those that long ago ceased their stellar activity, 
likely the massive ones. 
Spiral galaxies occupy  nearly the same location on the MR-plane thus suggesting that their ongoing
star formation is not affecting the overall situation of mechanical equilibrium of the whole system.  Given
that linear relations have been used (in compliance with the \textit{euclidean ``ruler and compass'' method}),  
the result is a straight line visualized by the large empty squares.   The
correct evaluation would by given by  numerically folding the adopted HGF, \citet{Lukic2007} in our case, with the loci 
on the MR-plane populated at the present time by galaxies of different mass but the same initial density born at 
different redshifts. The result is the  crossed dark-red line
shown in Fig. \ref{mass_radius}, i.e. eqs. (\ref{shepherd_rel}) and/or (\ref{good_relation}). 

(ix) Finally, it is worth noting that the slope of MRR derived from the HGF is about 1 in the range of massive galaxies 
(say above $10^{12}\, M_\odot$), i.e. formally identical to the MRR that one would derive
from the  virial theorem.  This coincidence might  suggest that the observed MRR and its slope are driven by 
the virial condition whereas
the true driver of the MRR slope is the HGF, more precisely its fall off toward high values of the halos' masses at any
value of the redshift. On the other hand, all objects along the MRR are indeed in virial conditions once any mechanical
process and star  formation are at rest.

\section{ General remarks and conclusions}\label{conclusions}

 What can we learn from all these observational hints, and striking coincidence between data and theory?
Seeking for a coherent picture,  one is tempted to suggest the following scenario:

(i) According to their initial density galaxies of given total mass will distribute in the MR plane along
lines of type  A and/or B and/or C. The initial density is not constant,  but
decreases as $z_{f}$ tends to zero. The permitted mass intervals for the total (and
baryonic) mass of the galaxies is not constant, but   increases  at decreasing
redshift. At redshifts of about 1 to 2, galaxies with total mass up to  a few $10^{12}\, M_{\odot} $ are in
place and their MRR will extend up to the range populated by the galaxies we see today.

(ii) The expectation is that galaxies of any mass and initial density  crowd a strip bounded by the lines of
maximum and minimum initial density (say redshifts from about 10 to 1--2) and the line corresponding to their
maximum mass in  the HGF for the same redshifts.

(iii) In this context, the semi-plane below  either relation  (\ref{delta_MR3}) in the simple description or
relation (\ref{good_relation}) based on the Cosmic Galaxy Shepherd and the HGF in turn, is void of galaxies, because 
they would be too massive at given initial density (redshift) to be compatible with  the maximum
mass of the HGF in general for all values of the redshift.

(iv) In the semi-plane above relation (\ref{delta_MR3}) and/or (\ref{good_relation}), the available data  are not 
statistically complete, so
that in the range of low mass galaxies we have only those of the Local Group (M32  and $\omega$Cen include).
Dwarf galaxies of large effective radius are simply not in the sample.
Observations of the Fornax cluster \citep{Im_etal_2001} have revealed  a  new type
of DGs. They have intrinsic sizes of about 100 pc and are more compact and
less luminous than other known compact dwarf galaxies, yet much brighter than globular clusters. Their
absolute magnitude is about $M_B=-13$ (two magnitudes fainter than M32). Are these objects
representing the low mass tail of galaxies falling along line A in Fig. \ref{mass_radius}? More observational 
data may clarify this issue.

(vi) GCs  are more difficult to discuss because, being those we have considered  bound to
the Milky Way, they have certainly suffered  many external (and internal) dynamical processes that may have
changed their present mass and radius.

(vii) Finally, there are  GCGs  to comment. Masses of their bulk stellar content and  half-mass radii 
of the ideal sphere in which stars are contained, are by still highly uncertain so that on the MR-plane they do not trace a sharp 
MRR but rather a  broad region. Part of the scatter is perhaps to GCGs still evolving towards the equilibrium state, part to 
observational uncertainties. Suffice here that the same theoretical MRR can pass trough all the regions of the MR-plane crowed by 
objects going from GCs to GCGs.

(viii) To conclude,  we endeavor to speculate on the origin of the present-day
MRRs and their differences  passing from GCs to DGs,  ETGs and GCGs. The focus goes to DGs and ETGs,  however  also GCs and 
GCGs are considered.   For all of them,  we suggest a common origin stemming from
the interplay among several processes, i.e. the initial density together with total mass that drive the star
formation history, the efficiency of energy feed-back in triggering  galactic winds that
affect the mass and the size of the objects, the merger history of a galaxy,
and the steady increase of the  maximum mass reachable by the HGF at decreasing redshift and or decreasing
mean density of the Universe. High density (low mass) objects are the first to form stars in the
remote past followed by objects of lower density (higher mass). The MRRs of eq.(\ref{RsMs_Bernardi}) or
(\ref{delta_MR3}) simply reflect the above interplay. Galaxies as massive as a typical
$10^{12}\, M_{\odot}$ object can be in place at redshifts from 2 to 1. After the first initial activity, 
passive evolution is expected. They  generate the MRR of
dynamically stable and passively evolving objects in virial conditions, the shape of which is driven by the
cosmological distribution (mass and relative number density) of the halos  hosting the visible galaxy. Since
the HGF is not a linear function, the slope of the MRR continuously varies from $\simeq 0.5$ to $\simeq 1$,
 Objects in similar physical
conditions (passive evolution and virial regime) but with lower mass (say up to $10^{10}\, M_\odot$), closely follow
the  MRR of the dissipation-less  case (slope of about 0.5 or so) and yet are in virial conditions (see GCs and DGs like M32 and
$\omega$Cen).
All other galaxies being still far from these ideal conditions because of active star formation, presence of galactic
winds, occurrence of repeated mergers, etc., strongly deviate from the ideal MRR of objects in mechanical 
and thermal equilibrium. The information for GCGs is still in progress so that no firm conclusions can be drawn, even though 
they seem to fit for this scenario.

\begin{acknowledgements}
      We like to thanks Prof. L. Secco and Dr. R. Caimmi for very useful discussions on many aspects of this study.
C.C. thanks the Department of Physics and Astronomy of the Padua University for the hospitality and
computational support.
\end{acknowledgements}

%
%
\bibliographystyle{aa}           
\bibliography{Mass_Radius_biblio}    

\begin{appendix}
 \noindent

\section{The stellar-to-halo mass relation}\label{Appendix_B}

The ratio $M_s/M_D$ (and its inverse $M_D/M_s$) are key quantities in the calculation of the (half-mass or 
effective) radius of the stellar component of a galaxy, see eq.(\ref{mr3}) of Sect. \ref{Theo_MRR}. The 
analysis  of the Illustris data and the inspection of Tables \ref{tab0}, \ref{tab1}, and  
\ref{tab2} in Appendix \ref{Appendix_C} show that the ratios $M_s/M_D$  and $M_D/M_s$ in turn depend on the 
total  mass of the galaxy and the redshift $z_f$ at which the  bulk of star formation occurs. 

\begin{figure}
\centerline{
\includegraphics[width=8.0cm,height=8.0cm]{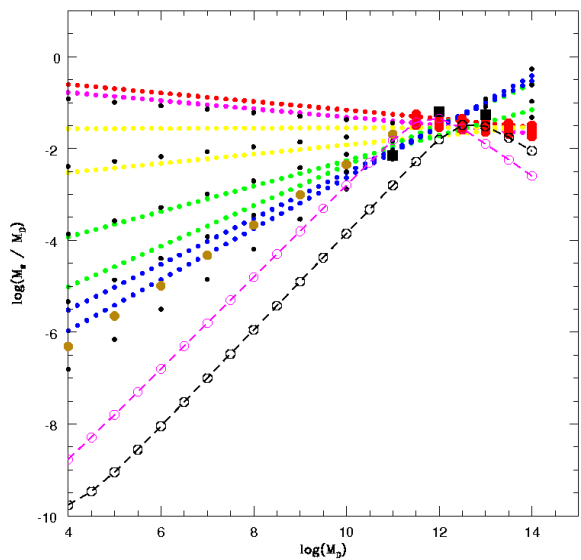}  }
\caption{ The relations $M_s/M_D$  $M_D$  at different redshifts for theoretical models. The dotted 
lines  of different colors
correspond to the eight values of the redshift z=0 and z=0.2 (top, red), z=0.6, z=1.0 (intermediate,
yellow), z=1.6 and z=2.2 (intermediate, green), z=3 and  to z=4 (bottom, blue). The black dots are the values
predicted by eq. (\ref{ratio_ms_md_eq}) at varying $\log M_D$ from 4 to 14 in steps of 1 and redshift $z$ from 0 
to 4 in steps of 1,
respectively. The large red and golden circles are the combination of eq. (\ref{md_ms_shan}) and
eq. (\ref{md_ms_fan}), respectively.  
 The open magenta (z=0) and black (z=3.95) circles are relation  $M_s/M_D$  vs $M_D$ at different redshifts according to 
\citet{Girelli_etal_2020}.
Note how all 
relations agree each other at $log M_D \simeq 12$ whereas they badly disagree each other at lower values of $M_D$. 
All masses 
are in solar units. }
\label{md_ms_zeta}
\end{figure}

 The dependence of the ratio $M_s/M_D$ 
on $M_D$ and redshift $z$ for the Illustris data is shown in Fig. \ref{md_ms_zeta}.    For low values of the
redshift (say below 0.6), this ratio gently  decreases with the mass $M_{D}$ (low mass galaxies are
slightly more efficient in building their stellar content);  the opposite occurs for higher redshifts, where 
the ratio $M_s/M_D$ increases with $M_{D}$, i.e. the stellar mass built-up in low
mass galaxies is expected to be less than in the massive ones. 

With the aid of the theoretical data for $M_D$ and $M_s$ presented in Sect.\ref{Theo_MRR},  we derive an 
analytical expression interpolating the ratio  $M_s/M_D(M_D,z)$ as function of $M_D$ and $z$

\begin{equation}
\log \frac{M_s}{M_D}= [0.218\,z - 0.101] \, \log M_D + [0.169 \, z - 2.227 ]  
\label{ratio_ms_md_eq}
\end{equation}
where the halo mass goes from $10^4 M_\odot$ to $10^{14} M_\odot$ and the redshift from 0 to 4. 
 The ratios $M_s/M_D$ predicted by eq. (\ref{ratio_ms_md_eq}) are indicated by the small black dots  
 of Fig. \ref{md_ms_zeta}. The range of application of relation (\ref{ratio_ms_md_eq}) is $0 leq z \leq 4$.

Similar relationships for the ratio $m= M_D/M_s$ can be found in literature, see for 
instance \citet{Fan_etal_2010},  
\citet{Shankar_etal_2006}, and more recently \citet{Girelli_etal_2020}. 

For $M_D \geq 10^{11}\, M_\odot$ \citet{Fan_etal_2010} propose the relation

\begin{equation}
 m = \frac{M_D} {M_s}= 25 \left( \frac{M_D}{10^{12}} \right)^{0.1} \left( \frac{1+z}{4} \right)^{-0.25}
\label{md_ms_fan}
\end{equation}
from which we derive the ratio $M_s/M_D$ shown in Fig. \ref{md_ms_zeta} by the red circles. In practice there 
is no dependence on the redshift.

Remarkably, the 
\citet{Fan_etal_2010}
curve agrees with the one we have derived from the Illustris models for values of the redshift smaller than 
about 1.6 (the slope is nearly identical). 

\citet[][and references]{Shankar_etal_2006} present a detailed analysis of the dependence of $M_s$ on 
$M_D$. First, they suggest  that for  $M_D < 10^{11}\, M_\odot$  the relation should be
\begin{equation}
m= \frac{ M_D}{M_s}  = C\, M_D^{-2/3}
\label{md_ms_shan}
\end{equation}
with $C$ a suitable proportionality constant to be determined. By imposing equality between the values of
$m$ determined with the two above relationships at the transition mass $M_D \geq 10^{11}\, M_\odot$, the 
proportionality constant is $\log C = 9.044$. The ratios $M_D/M_s$ predicted by eq.(\ref{md_ms_shan}) are 
shown in Fig.\ref{md_ms_zeta} by the dark golden circles. It is worth noting that  the 
\citet{Shankar_etal_2006} relation agrees 
with that of the Illustris models for redshifts in the range from 2 to 4. 

Amazingly, it is worth noting that in Fig. \ref{md_ms_zeta} the linear extrapolation of   the 
\citet{Fan_etal_2010} relationship (red golden circles) 
to lower masses and the linear extrapolation of the  \citet{Shankar_etal_2006} curve (dark golden circles) 
to higher values of the mass   encompass the  
theoretical predictions from the Illustris models for all the values of the redshift.

Second, \citet{Shankar_etal_2006}  derive an analytical expression for the relation between  
$M_s$ and  $M_D$  
\begin{equation}
   M_s = 2.3 \times 10^{10} M_\odot \frac{    (M_D/3\times10^{11} M_\odot)^{3.1}  } 
                                         {1 + (M_D/3\times10^{11} M_\odot)^{2.2}  }
\label{md_ms_shan2}
\end{equation}

\noindent for $M_D \geq 10^{11} M_\odot$. In this relation there is no explicit dependence on the redshift. 
The ratios 
$M_s/M_D$ predicted by eq.(\ref{md_ms_shan2}) are shown in Fig. \ref{md_ms_zeta}  by the black filled 
squares. 
Eq. (\ref{md_ms_shan2}) predict ratios $m(M_D,z)$ that agree with those from eq.(\ref{ratio_ms_md_eq}) derived 
from  Illustris data, eq.(\ref{md_ms_fan}) from  \citet{Fan_etal_2010}, and eq.(\ref{md_ms_shan}) only in the 
 region around $\log(M_D) \simeq 12$ and $z \simeq 0$.

In a very recent study \citet{Girelli_etal_2020} have thoroughly investigated the  stellar-to-halo mass 
ratio of galaxies ($M_s/M_{DM} = 1/m$  in the mass interval $10^{11} < 
M_{DM} < 10^{15} $ and  redshifts from $z=0$  to $z=4$. They  use a statistical approach to link the 
observed galaxy stellar mass function on the COSMOS field  to the halo mass function from the 
$\Lambda$CDM-Dustgrain simulation  and derive an empirical model to describe the variation of the 
stellar-to-halo mass 
ratio as a function of the redshift. Finally they provide  analytical expressions for the function $M_s(M_D, 
z)$. The relations $M_s/M_D$ vs $M_D$ as function of the redshift obtained with the formalism of    
\citet{Girelli_etal_2020} are also shown in Fig. \ref{md_ms_zeta}
%
%
(the magenta and dark-olive-green dots joined by solid lines of 
the same color) and are compared with the results of our study. See also for a similar analysis the 
study of \citet{Engler_etal_2020}.

It is soon evident that while all studies agree on  the $M_s/M_D$ ratios for objects with halo mass in the interval 
$11.5 \leq log M_D \leq 12.5$ nearly independently of the redshift, they badly disagree each other going to lower values of 
the halo mass. Furthermore, they also disagree with the theoretical results predicted by Illustris.

To cope with this major points of uncertainty on the correct function $M_s(M_D, z)$, we prefer to provisionally 
simplify the 
whole issue  and, basing on the results from the few galaxy models to disposal (at redshift $z=0$ by construction),  
we derive  the following dependence
for the ratios $M_s/M_D$ and its inverse $m = M_D/M_s$
\begin{eqnarray}
\log \frac{M_s}{M_D} &=& -0.062\, \log M_D - 0.429  \nonumber \\
\log \frac{M_D}{M_s}  &=&  0.062\, \log M_D + 0.429 
\label{final_m}
\end{eqnarray}
that do  not depend on the redshift. However, since eq. (\ref{mr3}) in which the ratio 
$m(M_D,z)$ is used 
contains its own dependence of the redshift, the final results still depend on this important parameter.

\section{Hydrodynamical models of ETGs}\label{Appendix_C}

\subsection{The pure monolithic scheme}\label{App_A_Monolithic}

\citet{Chiosi_Carraro_2002}, by means of  hydrodynamical simulations, incorporating radiative cooling, star
formation, energy  feed-back and chemical evolution,   addressed the problem of the formation and evolution
of ETGs (from dwarf to normal/giant systems) in the framework of the  SCDM cosmological model of the Universe and
monolithic scheme of galaxy formation. The cosmological parameters were  $H_0=65\, km\,s^{-1}\,Mpc^{-1}$,
Baryonic to Dark Matter ratio 1 to 9, i.e. for $M_T= M_{BM}+M_{DM}$ $M_{BM} = 0.1 M_{T}$ and
$M_{DM} =0.9 M_{T}$, and the redshift
at which galaxies were supposed to start the collapse ($z_{f}=1$ and $z_{f}=5$). 
A lump of mass (perturbation) with spherical symmetry
whose density exceeds a critical value is let 
collapse toward the virial condition. A MonteCarlo procedure is adopted to fix the initial coordinates
and velocities of the DM and BM particles. Table \ref{tab0} summarizes the
key parameters and results of the models. The key results of this study are:
(i)   The duration, strength and shape
of the star formation rate as function of time
strongly depend on the galaxy mass and the initial density:
(a) Galaxies of high initial density   and total mass over an ample range undergo a prominent initial episode of
SF ever since followed by quiescence.
(b) The same applies to high mass galaxies of low initial density, whereas the low mass ones undergo
a series of burst-like episodes  that may stretch up to the present. The details of their SF history are also very
sensitive to the value of the initial density.

(ii) The  mass turned into stars per unit total mass of the galaxy is nearly  constant which means that the engine
at work is the same.

(iii) The ratio between the left-over gas to the initial total BM decreases at
increasing total mass of the galaxy. 

(iv) As a result of star formation  in the central core of the galaxy and
consequent gas heating, large amounts of gas are
 pushed to large distances. After cooling, part of the gas falls
 back towards the central furnace.

(v) Galactic winds  occur. In general all galaxies are able to eject
part of their gas content into the inter-galactic medium.
However, the percentage of the ejected material increases
at decreasing galaxy mass.

\begin{table*}
\tabcolsep 0.2truecm
\caption{Summary of the properties of the hydrodynamical  model galaxies by \citet{Chiosi_Carraro_2002}.
\textbf{Left--Initial conditions}: $M_T$ is the total mass of
the proto-galaxy,  $M_{DM}$ and $M_{BM}$ the mass of DM and
BM, respectively, $R_{200}$  the initial radius, $\langle\rho\rangle$ the mean mass density. Masses are in
$M_\odot$, radii in kpc, and densities in $M_{\odot}/kpc^3$. \textbf{Middle--Final results}:
the present day star $M_s$ and gas $M_g$ mass contents, the  half-mass radius
$R_{s}$ and central velocity dispersion  $\sigma_{s}$ of the stellar mass, and half-mass radius
$R_{DM}$ and central velocity dispersion  $\sigma_{DM}$ of DM.
\textbf{Right--Galactic winds}: $M_{g,esc}$ amount of gas inside the
radius at which the radial velocity is equal to the escape velocity;
 $\Delta M_{g,w}$ amount of gas lost in the galactic wind;  $\Delta M_{g,w} / M_g$
percentage of gas lost in the wind with respect to the amount left
over by Star Formation.  Masses are in $M_\odot$, radii in kpc, densities in $M_\odot /kpc^3$, and
velocity dispersions in km/s}
\begin{tabular}{|c lll c l|ll llll| lll|}
\hline
\multicolumn{1}{|c}{$Model$}   &
\multicolumn{1}{c}{$M_{T}$}    &
\multicolumn{1}{c}{$M_{DM}$}   &
\multicolumn{1}{c}{$M_{BM}$}   &
\multicolumn{1}{c}{$R_{200}$}  &
\multicolumn{1}{c|}{$\langle \rho \rangle_0$} &
\multicolumn{1}{c}{$M_{s}$}    &
\multicolumn{1}{c}{$M_{g}$}    &
\multicolumn{1}{c}{$R_{s}$}    &
\multicolumn{1}{c}{$\sigma_{s}$}  &
\multicolumn{1}{c}{$R_{DM}$}   &
\multicolumn{1}{c|}{$\sigma_{DM}$}  &
\multicolumn{1}{c}{$M_{g,esc}$} &
\multicolumn{1}{c}{$\Delta M_{g,w}$}&
\multicolumn{1}{c|}{ ${ \frac{\Delta M_{g,w}}  {M_g }}$  } \\
\hline
\multicolumn{15}{|c|}{Models with mean initial density
   $\langle \rho\rangle \simeq 200\times \rho_u(z)$  with $z\simeq 5$}  \\
\hline
(1A)    & 1e13  &  9.0e12 &1.0e12 &83   & 4e6 & 6.4e11  & 3.6e11 & 1.87   &  530 &  25  & 646 & 1.5e{11}  & 2.1e{11} & 58  \\
(6A)    & 1e9   &  9.0e8  &1.0e8  &4    & 4e6 & 3.2e7   & 6.8e7  & 0.07   &   17 &   1  & 20  & 2.4e7     & 4.5e7    & 66  \\
\hline
\multicolumn{15}{|c|}{Models with mean initial density $\langle\rho\rangle \simeq  5\times \rho_u(z)$ with
$z \simeq 1$}  \\
\hline
(1B)    & 5e13   & 4.5e13 &5.0e12 & 681&  4e4 & 4.4e12  & 0.6e12 & 29     &  365 & 276 & 654 & 4.0e{11}  &2.0e{11}  &33  \\
(2B)    & 5e12   & 4.5e12 &5.0e11 & 316&  4e4 & 4.0e11  & 1.0e11 & 16     &  215 & 170 & 324 & 6.5e{10}  &3.5e{10}  &35  \\
(3B)    & 1e12   & 9.0e11 &1.0e11 & 184&  4e4 & 6.4e10  & 3.6e10 &  9     &   98 &  63 & 153 & 1.9e{10}  &1.7e{10}  &47  \\
(4B)    & 2e11   & 1.8e11 &2.0e10 & 108&  4e4 & 1.2e10  & 0.8e10 &  5     &   69 &  53 & 119 & 0.4e{10}  &0.4e{10}  &50  \\
(5B)    & 1e10   & 9.0e9  &1.0e9  &  40&  4e4 & 4.0e8   & 6.0e8  &  2     &   33 &  19 &  55 & 2.5e{8}   &3.5e{8}   &58  \\
(6B)-LD & 1e9    & 9.0e8  &1.0e8  &  35&  5e3 & 1.3e7   & 8.7e7  &  9     &    3 &  1  &  13 & 1.5e{7}   &0.8e7     &35  \\
(6B)-ID & 1e9    & 9.0e8  &1.0e8  &  19&  4e4 & 2.5e7   & 7.5e7  &  1     &    6 &  10 &  15 & 3.0e{7}   &4.5e7     &60  \\
(6B)-HD & 1e9    & 9.0e8  &1.0e8  &  16&  6e4 & 7.7e7   & 2.3e7  &0.3     &    5 &  19 &  19 & 6.4e{7}   &2.3e7     &24  \\
(7B)    & 1e8    & 9.0e7  &1.0e7  &   9&  4e4 & 1.8e6   & 8.2e6  &  1     &    3 &   4 &  10 & 3.0e{6}   &5.1e6     &63  \\
\hline
\end{tabular}
\label{tab0}
\end{table*}

\begin{table*}
\caption{The early-hierarchical galaxy models by \citet{Merlin2010,Merlin2012}. 
Left to right: total initial mass $M_T=M_{DM} + M_{BM} $ 
[in units of $10^{12} M_{\odot}$], corresponding (initial) gas mass $M_g$ [in units of $10^{12} M_{\odot}$], 
redshift $z_i$ at which the proto-halo is selected from the background in the cosmological grid, 
mean halo over-densities $(<\rho> -\rho_b)/\rho_b = [\delta\rho-1]_{i}$ ($\rho_b$ is the background density) at the redshift $z_i$, 
initial proper physical radius of the halo [in kpc], redshift of the last computed model $z_f$, 
corresponding age $t_f$ [in Gyr], virial radius of the whole system $R_{vir}$ [in kpc], half-mass  
radius $R_{1/2}$ at $z_f$, and the total stellar mass $M_s$ at the final redshift [in units of $10^{12}M_{\odot}$]. } 
\centering
\begin{tabular}{|l| l|l|l|c|l|l|l|l|l|l|}
\hline
Model & $M_{T}$  & $M_{g,i}$ & $z_{i}$ & $[\delta\rho-1]_{i}$ &  $R_{i}$ & $z_{f}$ & $t_{f}$ & $R_{vir}$ & $R_{1/2}$ & $M_{s}$  \\
\hline
HDHM  & 17.5    & 2.90     & 46 & 0.12  & 97.17  & 0.22  & 11.0  & 153.0  & 15.6 & 0.75  \\
\hline
MDHM  & 17.5    & 2.90     & 39 & 0.12  & 114.31 & 0.77  & 8.0   & 141.8  & 15.2 & 0.74 \\
\hline
LDHM  & 17.5    & 2.90     & 33 & 0.12  & 134.49 & 0.50  & 8.7   & 133.8  & 14.1 & 0.73 \\
\hline
VLDHM & 17.5    & 2.90     & 23 & 0.12  & 194.34 & 0.83  & 6.6   & 112.5  & 10.8 & 0.63 \\
\hline
HDMM  & 0.269   & 0.0445   & 54 & 0.11  & 20.99  & 1.0   & 5.8   & 37.6   & 5.5  & 0.020 \\
\hline
MDMM  & 0.269   & 0.0445   & 45 & 0.11  & 24.69  & 0.75  & 7.0   & 35.7   & 5.4  & 0.019 \\
\hline
LDMM  & 0.269   & 0.0445   & 38 & 0.11  & 29.05  & 0.58  & 8.1   & 33.3   & 4.8  & 0.019 \\
\hline
VLDMM & 0.269   & 0.0445   & 26 & 0.11  & 41.98  & 0.15  & 11.8  & 28.3   & 4.7  & 0.017 \\
\hline
HDLM  & 0.00417 & 0.000691 & 63 & 0.09  & 4.48   & 0.36  & 9.7   & 9.2    & 2.3  & 0.00015 \\
\hline
MDLM  & 0.00417 & 0.000691 & 53 & 0.09  & 5.27   & 0.22  & 11.0  & 10.0   & 2.1  & 0.00014 \\
\hline
LDLM  & 0.00417 & 0.000691 & 45 & 0.09  & 6.20   & 0.05  & 13.0  & 11.8   & 2.0  & 0.00014 \\
\hline
VLDLM & 0.00417 & 0.000691 & 31 & 0.09  & 8.96   & 0.0   & 13.7  & 10.5   & 2.5  & 0.00010 \\
\hline
\end{tabular}
\label{tab1}
\end{table*}

\subsection{The hierarchical scheme}

\citet{Merlin2010,Merlin2012} with the aid of the  parallel hydrodynamical code EvoL  produced a number of 
models for
ETGs with different mass and/or initial density (twelve cases in total). 

The models were followed from the epoch
of their detachment from the linear regime, i.e. $z_{i} \geq 20$, to a final epoch (redshift $z_f$) varying
from model to model (however with $z_{f}\leq 1$). The simulations included radiative cooling down to 10 K, star formation,
stellar energy feedback,  re-ionizing photo-heating
background, and chemical enrichment of the interstellar medium. The reader can refer to the cited papers for all the
details on the method and the code in use as well as on model results.

The assumed cosmology was the standard $\Lambda-$CDM, with $H_0$=70.1 km/s/Mpc, flat geometry,
$\Omega_{\Lambda}$=0.721, $\sigma_8$=0.817; the baryonic fraction $\simeq 0.1656$.

Each model starts as a sphere cut at certain redshift $z_i$ from a wider cosmological simulation in which  density
fluctuations exist. At the center  of  the sphere there is a peak of given density contrast with respect to the cosmological
background. The density contrast  is measured by the quantity $[\delta \rho -1] = [(<\rho> - \rho_b)/\rho_b] $, 
where $<\rho>$ is the mean density of
the perturbation and $\rho_b$ the density of the background. The central density peak has a given total mass $M_T$, 
sum of the DM and BM components. This is the proto-halo of our model galaxy in which stars will be
formed at later times. The cosmological simulation provides the initial positions and velocities of all the particles
in the proto-halo.  An outward radial component is added to the velocity of each particle to take the expansion of the
Universe into account.  The initial conditions are set in such a way  that each model is a re-scaled
version of a single reference proto-galactic halo, with different total mass and/or initial over-density. 
The proto-galactic halos are then followed through their early stages of expansion following the Hubble flow,
the turn around, and the collapse. The redshift at which the collapse begins varies from model to model and inside the
same model from the center to the outer regions. In general the collapse occurs during the redshift interval
$4 > z > 2 $, it starts first in the central regions and gradually moves outwards. The collapse is complete at
redshift $z\simeq  2$.
All the models develop a central stellar system, with a spheroidal shape. 
Massive halos ($M_{T}\simeq 10^{13} M_{\odot}$) experience a single, intense burst of star formation
(with rates $\geq 10^3 M_{\odot}$/yr) at early epochs, consistently with observations, with a less pronounced
dependence on the initial over-density; intermediate mass halos ($M_{T}\simeq 10^{11} M_{\odot}$) have star
formation histories that strongly depend on their initial over-density, i.e. from a single peaked to a long lasting
period of period of activity with
strong fluctuations in the rate; finally small mass halos ($M_{T}\simeq 10^{9} M_{\odot}$) always have fragmented
histories, resulting in multiple stellar populations, due to the so-called ``galactic breathing'' phenomenon.
They confirm the correlation between the
initial properties of the proto-halos and their star formation histories already found by
\citet{Chiosi_Carraro_2002}.
The models have morphological, structural and photometric properties comparable to real galaxies, in general
closely
matching the observed data; there are minor discrepancies that are likely of numerical origin
\citep[see][\, for all other details]{Merlin2012}.
These models can be classified as \textit{early hierarchical} because they undergo repeated episodes of mass
accretion of sub-lumps of matter inside the original density contrast in very early epochs and essentially 
evolve in isolation ever since. 
These reference models are calculated adopting a star formation efficiency $\epsilon_{sf} = 1$. This value is  
larger than current estimates from observational data in the local Universe, i.e. $\epsilon_{sf} \simeq 0.025$
\citep{Lada2003, Krumholz2007}, and theoretical considerations on the global star-to-total mass ratio in 
galaxies that
suggest $\epsilon_{sf} \leq 0.1$. However, adopting a high value of $\epsilon_{sf}$ allows for a strong 
reduction of
the computational time, while preserving the basic properties of the models
\citep[for a complete discussion see][]{Merlin2012}.
The properties of all computed model galaxies are listed in Table \ref{tab1}.

\begin{table*}
\caption{The ancillary models of \citet{Chiosi_etal_2012}. The meaning of the symbols is as follows: 
Model is the two-letter string identifying the 
model according to the mass: MM for intermediate mass galaxy $2.69 \times 10^{11}M_\odot$ and LM for the low mass case
$4.17 \times 10^{9}M_\odot$;  $f_\delta$ is the multiplicative factor of the initial over-density of  
the reference model of the same mass; $\epsilon_{sf}$ is the dimensionless efficiency of the star
formation rate, the symbol $\epsilon_{Z}$ means that the efficiency is supposed to increase from $\epsilon_{sf}=0.1$ for 
$Z$=0.0001 to $\epsilon_{sf} = 1$ for $Z$=0.01 (close to the solar value). All other symbols have the same meaning as in 
Table \ref{tab1}.  } \centering
\begin{tabular}{|r|r|l |l|l|r  |l|l|l |l|l|l |l|}
\hline
Model &$f_\delta$ &$\epsilon_{sf}$   &$M_{t}$& $M_{g,i}$  & $z_{i}$      & $[\delta\rho-1]_{i}$ & $R_{i}$ & $z_{f}$ & $t_{f}$ & $R_{vir}$ & $R_{1/2}$ & $M_{s}$  \\
\hline
MM    &  20     &    1    & 0.269 & 0.0445     & 181 & 0.18 & 6.3  & 62.0  & 0.04   &  1.8   & 0.63  & 0.010\\
\hline
MM    &  20     &    1    & 0.269 & 0.0445     & 181 & 0.18 & 6.3  & 30.4  & 0.10   &  4.9   & 1.26  & 0.054 \\
\hline
MM    &  20     &    0.1  & 0.269 & 0.0445     & 181 & 0.18 & 6.3  & 56.0   & 0.04  &  2.1   & 0.08  & 0.0050 \\
\hline
MM    &  20     &    0.1  & 0.269 & 0.0445     & 181 & 0.18 & 6.3  & 49.0  & 0.05   &  2.6   & 0.08  & 0.0074  \\
\hline
MM    &  20     &    $\epsilon_{Z}$  & 0.269 & 0.0445     & 181 & 0.18 & 6.3  & 59.0  & 0.04   &  2.0   & 0.07  & 0.0036 \\
\hline
MM    &  20     &    $\epsilon_{Z}$  & 0.269 & 0.0445     & 181 & 0.18 & 6.3  & 46.0  & 0.05   &  2.9   & 0.09  & 0.0083 \\
\hline
MM    &  15     &    1    & 0.269 & 0.0445     & 140 & 0.18 & 8.2  & 30.4  & 0.10   &  4.5   & 1.08  & 0.026 \\
\hline
MM    &  12     &    1    & 0.269 & 0.0445     & 117 & 0.17 & 9.8  & 30.4  & 0.10   &  4.1   & 0.92  & 0.011 \\
\hline
MM    &  12     &    1    & 0.269 & 0.0445     & 117 & 0.17 & 9.8  & 18.8  & 0.20   &  7.4   & 0.98  & 0.015 \\
\hline
MM    &  12     & $\epsilon_{Z}$  & 0.269 & 0.0445     & 117 & 0.17 & 9.8  & 36.0  & 0.08   &  3.1   & 0.07  & 0.0032 \\
\hline
MM    &  12     & $\epsilon_{Z}$  & 0.269 & 0.0445     & 117 & 0.17 & 9.8  & 30.4  & 0.10   &  4.2   & 0.10  & 0.00575 \\
\hline
MM    &  5      &    1    & 0.269 & 0.0445     & 68  & 0.14 & 16.7 & 7.0   & 0.70   &  19.2  & 5.75  & 0.019 \\
\hline
MM    &  5      &    1    & 0.269 & 0.0445     & 68  & 0.14 & 16.7 & 4.1   & 1.50   &  34.3  & 6.76  & 0.057 \\
\hline
LM    &  20     &    1    & 0.00417 & 0.000691 & 112 & 0.16 & 2.6  & 25.3  & 0.13   &  1.2   & 0.32  & 0.00016 \\
\hline
LM    & 20      &    1    & 0.00417 & 0.000691 & 112 & 0.16 & 2.6  & 18.8  & 0.20   &  1.8   & 0.31  & 0.00017 \\
\hline
LM    & 20      &    0.1  & 0.00417 & 0.000691 & 112 & 0.16 & 2.6  &  25.4  & 0.13  &  1.3   & 0.02  & 0.000091 \\
\hline
\end{tabular}
\label{tab2}
\end{table*}

Using the same numerical code and approach for the initial conditions, \citet{Chiosi_etal_2012} 
produce a
group of \textsf{ancillary models}, whose initial parameters and key results are listed in Table \ref{tab2}. 
These models
are calculated to explore the consequences of much higher initial density contrasts and/or lower star
formation efficiencies
$\epsilon_{sf}$.  The the effect of the higher initial density is already known from the old calculations
\citet{Chiosi_Carraro_2002}
and the models by \citet{Merlin2012}: galaxies of the same mass will be shifted on the MR-plane to smaller 
radii.

As expected by decreasing the efficiency
$\epsilon_{sf}$, star formation is delayed or even inhibited. The gas continues to flow into the gravitational
potential
well till the threshold density for star formation to occur is reached and /or  sufficient number of stars are
formed,
the newly born galaxy has much smaller dimension with respect to the corresponding object with higher 
efficiency of star formation.  

All the ancillary models  are calculated limited to the  very early 
evolutionary stages, to show in the MR-plane the initial position of a model galaxy in which the gas content 
has reached densities much higher than the formal mean background density fixed by cosmology. The parameter  
$f_\delta$ indicates the factor by which the initial density of same model in the first group (Table 
\ref{tab1}) is scaled. As already said in these models we also change the efficiency of star formation 
$\epsilon_{sf}$ as indicated in column (3) Table \ref{tab2}. In a few models, indicated by $\epsilon_{Z}$, 
the efficiency of star formation increases with metallicity $Z$. The efficiency goes from 
$\epsilon_{sf}=0.1$ for $Z$=0.0001 to $\epsilon_{sf} = 1$ for $Z$=0.01
(close to the solar value). No special meaning must be given to this relation, it is simply meant to evaluate 
the effect of a star formation efficiency increasing with the metallicity. In any case, this effect plays 
a marginal role on the position of  the model galaxies on the MR-plane, see the entries of Table \ref{tab2}.

\end{appendix}

\end{document}